\documentclass{article}
\usepackage{arxiv}
\usepackage[utf8]{inputenc} 
\usepackage[T1]{fontenc}    
\usepackage{hyperref}       
\usepackage{url}            
\usepackage{booktabs}       
\usepackage{amsfonts}       
\usepackage{nicefrac}       
\usepackage{microtype}      
\usepackage{lipsum}         
\usepackage{graphicx}
\usepackage[round]{natbib}
\usepackage{doi}
\usepackage{xcolor}
\usepackage{comment}
\usepackage{amsmath}
\usepackage{booktabs}
\usepackage{cleveref}       

\title{Characterization Of Diseases In Temporal Comorbidity Networks}

\newif\ifuniqueAffiliation

\uniqueAffiliationtrue

\ifuniqueAffiliation 

\usepackage{authblk}

\newbox{\orcid}\sbox{\orcid}{\includegraphics[scale=0.06]{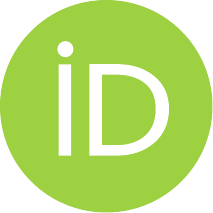}} 
\author[1,2]{%
	\href{https://orcid.org/0009-0003-8602-3552}{\usebox{\orcid}\hspace{1mm} Yuri Gardinazzi }%
}
\author[3]{%
	\href{https://orcid.org/0009-0009-3551-1680}{\usebox{\orcid}\hspace{1mm} Roger Gonzaléz March}
}

\author[4]{%
	\href{https://orcid.org/0000-0002-2757-1896}{\usebox{\orcid}\hspace{1mm} Suprabhath Kalahasti}%
}

\author[6]{%
	\href{https://orcid.org/0009-0005-9313-1708}{\usebox{\orcid}\hspace{1mm} Andrea  Montaño Ramirez}%
}
\author[7]{%
	\href{https://orcid.org/0000-0000-0000-0000}{\usebox{\orcid}\hspace{1mm} Matteo Neri}%
}

\author[9]{%
	\href{https://orcid.org/0009-0006-9828-257X}{\usebox{\orcid}\hspace{1mm} Cicely Nguyen }}

\author[10,11]{%
	\href{https://orcid.org/0009-0006-6439-9243}{\usebox{\orcid}\hspace{1mm} Giovanni Palermo }%
}

\author[12]{%
	\href{https://orcid.org/0000-0000-0000-0000}{\usebox{\orcid}\hspace{1mm} Erik Weis}%
}
   
\author[13,14,15]{%
	\href{https://orcid.org/0000-0001-9923-1172}{\usebox{\orcid}\hspace{1mm} Katharina Ledebur }%
}

\author[13,14,15]{%
	\href{https://orcid.org/0000-0001-7168-3310}{\usebox{\orcid}\hspace{1mm} Elma Dervić\thanks{
    \texttt{yuri.gardinazzi@areasciencepark.it},
    \texttt{roger.gonzalez@bsc.es}, 
    \texttt{suprabhath.kalahasti@iplesp.upmc.fr}, 
    \texttt{amontanoramirez2@ucmerced.edu},  
    \texttt{matteo.neri@etu.univ-amu.fr}, 
    \texttt{cicely.nguyen@stud.uni-heidelberg.de},  \texttt{giovanni.palermo@cref.it}, \texttt{weis.er@northeastern.edu},  \texttt{ledebur@csh.ac.at}, \texttt{dervic@csh.ac.at}}}%
}

\affil[1]{AREA Science Park,Trieste, Italy}
\affil[2]{University of Trieste, Trieste, Italy}
\affil[3]{Barcelona Supercomputing Center, Barcelona, Spain}
\affil[4]{Sorbonne Université, INSERM, Institut Pierre Louis d'Epidémiologie et de Santé Publique, Paris, France}
\affil[6]{University of California Merced, California, USA}
\affil[7]{Aix-Marseille Université, CNRS, Institut de Neurosciences de la Timone, Marseille, France}
\affil[9]{Heidelberg University, Heidelberg, Germany}
\affil[10]{Sapienza University of Rome, Rome, Italy}
\affil[11]{Enrico Fermi Research Center, Rome, Italy}
\affil[12]{Network Science Institute, Northeastern University, Boston, USA}
\affil[13]{ Institute of the Science of Complex Systems, Center for Medical Data Science, Medical University of Vienna, Vienna, Austria}
\affil[14]{ Complexity Science Hub, Vienna, Austria} 
\affil[15]{Supply
Chain Intelligence Institute Austria (ASCII), Vienna, Austria}

\hypersetup{
pdftitle={A template for the arxiv style},
pdfsubject={q-bio.NC, q-bio.QM},
pdfauthor={David S.~Hippocampus, Elias D.~Striatum},
pdfkeywords={First keyword, Second keyword, More},
}

\begin{document}
\maketitle

\begin{abstract}

Comorbidity networks, which capture disease–disease co-occurrence usually based on electronic health records, reveal structured patterns in how diseases cluster and progress across individuals. However, how these networks evolve across different age groups and how this evolution relates to properties like disease prevalence and mortality remains understudied. To address these issues, we used publicly available comorbidity networks extracted from a comprehensive dataset of 45 million Austrian hospital stays from 1997 to 2014, covering 8.9 million patients. These networks grow and become denser with age. We identified groups of diseases that exhibit similar patterns of structural centrality throughout the lifespan, revealing three dominant age-related components with peaks in early childhood, midlife, and late life. To uncover the drivers of this structural change, we examined the relationship between prevalence and degree. This allowed us to identify conditions that were disproportionately connected to other diseases. Using betweenness centrality in combination with mortality data, we further identified high-mortality bridging diseases.

Several diseases show high connectivity relative to their prevalence, such as iron deficiency anemia (D50) in children, nicotine dependence (F17), and lipoprotein metabolism disorders (E78) in adults. We also highlight structurally central diseases with high mortality that emerge at different life stages, including cancers (C group), liver cirrhosis (K74), subarachnoid hemorrhage (I60), and chronic kidney disease (N18). These findings underscore the importance of targeting age-specific, network-central conditions with high mortality for prevention and integrated care.

\end{abstract}


\section{Introduction}

Significant advances in medical and health innovations and solutions have extended the lifespan of populations around the world. On average, people today live longer and healthier lives. Factors that have contributed to this are improved access to education, diet and non-sedentary activities along with improvements in medications, diagnostics and screening \citep{naghavi2024global, xia2024global, steel2025changing, partridgeFacingGlobalChallenges2018a}. 

A consequence of this trend along with drop in fertility rates is an ageing population~\citep{AgeingHealtha}. In the 21st century, population ageing is expected to pose challenges to the healthcare system in terms of the burden (economic and social) they present. Public health frameworks that address these needs have to be developed to prevent healthcare shocks and develop long term resilience \citep{sanderChallengesHumanPopulation2015,europeancommission.statisticalofficeoftheeuropeanunion.AgeingEuropeLooking2020,thaneGrowingBurdenAgeing1987,beardWorldReportAgeing2016}. It is also important to note that improvements in longevity do not necessarily mean healthy, disease free lives. Delaying the aging process with improvements in quality of life are pressing research needs \citep{crimminsLifespanHealthspanPresent2015}. 

The disease burden does not simply accumulate as populations age. As populations age, the complexity of disease trajectories increases, often involving interactions between conditions that span physiological systems. Chronic conditions increasingly interact across physiological systems, forming entangled patterns of multimorbidity. These interactions can amplify vulnerability and accelerate health decline. Mortality does not result from isolated conditions, but rather from the structure of disease combinations and their connectivity \citep{janiRelationshipMultimorbidityDemographic2019, menottiPrevalenceMorbidityMultimorbidity2001}. Identifying the structural role of specific diseases in these evolving networks may uncover key points for (early) preventive intervention.

By identifying such central and bridging diagnoses across the life course a network-based view can deepen this understanding by highlighting conditions linked to increased mortality, prolonged hospital stays, higher readmission rates, and overall healthcare burden for each phase of life \citep{hidalgo2009dynamic}. Comorbidity networks provide a powerful framework to enable such analyses at scale.

In these networks, nodes represent diagnoses, and edges indicate statistically significant co-occurrence relationships based on large-scale health records \citep{siggaard_disease_2020, roque_using_2011, aguado_morbinet_2020, bao_exploring_2023, huNetworkBiologyConcepts2016}. Various statistical methods can be used to construct these networks from raw data \citep{mConstructingDiseaseComorbidity2023,zhaoComorbidityNetworkAnalysis2023,fotouhiStatisticalMethodsConstructing2018,divoCOPDComorbiditiesNetwork2015}. Comorbidity networks have been used to map disease burdens, identify high-risk clusters, and predict health outcomes \citep{cramerComorbidityNetworkPerspective2010, cruz-avilaComorbidityNetworksCardiovascular2020, capobiancoComorbidityNetworksDisease2015}. However, most of this work has been cross-sectional, lacking a temporal view of how these networks evolve over the life course \citep{siahMultimorbidityMortalityData2022}.

Based on a recent dataset published by \citealp{dervic_comorbidity_2025}, which provides a detailed view of comorbidity patterns among the Austrian population, derived from 45 million hospital stays of 8.9 million patients between 1997 and 2014, we present a thorough analysis of how comorbidity networks change across different age groups. Previous research on age-stratified comorbidity networks has shown that disease clusters are relatively distinct in early life but become increasingly interconnected with age \citep{chmiel_spreading_2014}.

This structural transformation has been described by \citealp{dervic_comorbidity_2025,chmiel_spreading_2014,siggaard_disease_2020}, but its deeper implications are unclear. Does the increasing complexity of comorbidity networks with age reflect more fundamental shifts in disease organization, or is it simply a consequence of rising disease prevalence and a few highly connected conditions? To address this question, we analyze the evolving architecture of age-stratified comorbidity networks. 

As a first step to understand structural changes in comorbidity networks throughout the life course, we use non-negative matrix factorization (NMF) \citep{sra_generalized_2005} to identify distinct temporal patterns in disease centrality and trace which disease groups become more prominent with age.

Building on this temporal decomposition, we then examine the relationship between disease prevalence and network connectivity over age groups to identify potential structural drivers of network entanglement. Finally, we analyze bridging diseases, which connect otherwise separated parts of the network, and evaluate their mortality burden. This enables us to pinpoint structural pathways to severe outcomes and identify candidates for prevention and early intervention.

Comorbidity networks offer a powerful framework for disentangling how individual diseases condition on one another and contribute to lifetime outcomes such as mortality. This systems-level perspective holds promise for advancing population-level personalized medicine \citealp{haugHighriskMultimorbidityPatterns2020,passarelli-araujoMachineLearningComorbidity2022,koskinenDatadrivenComorbidityAnalysis2022}. Despite the widespread availability of electronic health records, the role of comorbidity network structure in shaping patient mortality remains an underexplored area of research \citep{forteComorbiditiesMedicalHistory2019}.

\section{Methods} \label{sec:Method}
\subsection*{Dataset}
We used a comprehensive dataset on all hospital records in Austria from 1997 to 2014, which was provided by the Austrian Ministry of Health \citep{dervic_comorbidity_2025}. The dataset includes approximately 45 million hospital stays for over 9 million individuals. Each record contains the following information: patient sex; 5-year age group; length of stay; discharge reason (e.g., discharge, transfer, or death); and primary and secondary diagnoses, which are coded at the 3-digit ICD-10 level \citep{Icd10}. 

In the following, we present the methods used on this dataset of networks provided in \citealp{dervic_comorbidity_2025} and in-hospital mortality rates (defined as the percentage of patients diagnosed with the diagnosis in a specific age group who die in-hospital) per diagnoses, sex, and age group used in \citealp{dervic2024unraveling}. 

The networks in the dataset represent comorbidity relationships between diseases. Each node corresponds to a disease, and edges connect diseases that have been observed together in patients. The edge weights represent the odds ratio between the connected diseases, weighted across years and filtered to retain only statistically significant connections (see the original paper \citealp{dervic_comorbidity_2025} for further details).
Thus, edges can be interpreted as links between diseases that co-occur more frequently than expected by chance.
We filter for nodes with at least one neighbor, discarding the ones having no significant comorbidities.
As for the disease prevalence and mortality, we consider the most recent values in the dataset (2014).

The networks analyzed are stratified by patient sex (\texttt{Male}, \texttt{Female}) and age group, in 10-year intervals from \texttt{0-9} to \texttt{70-79}, Table \ref{tab:graph-properties}.

\begin{table}[h]
    \centering
    \resizebox{16.5cm}{!}{%
       \begin{tabular}{
    l 
    *{8}{c} 
    @{\hspace{1em}} 
    *{8}{c} 
  }
  \toprule
  & \multicolumn{8}{c}{\textbf{Female}} 
  & \multicolumn{8}{c}{\textbf{Male}} \\ 
  \cmidrule(lr){2-9} 
  \cmidrule(lr){10-17} 

  \textbf{Age Group} 
  & 0-9 & 10-19 & 20-29 & 30-39 & 40-49 & 50-59 & 60-69 & 70-79 
  & 0-9 & 10-19 & 20-29 & 30-39 & 40-49 & 50-59 & 60-69 & 70-79 \\ 
  \midrule
\# Connected Nodes&111&109&152&200&294&333&364&384&136&99&132&208&310&366&374&359\\
Degree & 5.00 &2.85&4.57&6.07&9.20&12.07&14.52&21.78&6.25&2.41&4.73&6.07&9.61&13.41&17.23&20.39\\
Average Path &3.56&3.10&3.49&3.61&2.99&2.64&2.55&2.45&3.24&2.83&3.11&3.42&2.94&2.66&2.55&2.40\\
Betweenness & 14.57&6.57&18.40&45.67&78.01&83.70&95.37&99.79&18.71&3.64&10.24&40.29&86.38&101.44&99.41&85.54\\
Closeness & 0.30&0.39&0.38&0.29&0.34&0.38&0.39&0.41&0.34&0.55&0.43&0.35&0.34&0.38&0.41&0.41\\
Modularity & 0.49&0.65&0.59&0.51&0.40&0.35&0.29&0.24&0.43&0.72&0.54&0.51&0.41&0.34&0.28&0.25\\
  \bottomrule
  \end{tabular}
  }
    \caption{Summary of network properties across age groups in female and male comorbidity networks.}
    \label{tab:graph-properties}
\end{table}

\subsection*{NMF of Node Strength Centrality Evolution}

We first computed the node strength centrality for each
diagnosis within comorbidity networks stratified by age group, to examine how the central role of diagnoses changes over time. This resulted in a time series of centrality values for each node. We then calculated the pairwise correlation matrix across these centrality vectors to assess the similarity in their temporal trajectories. To identify distinct patterns of co-evolution, we applied \textit{K}-means clustering to the correlation matrix—a standard method for partitioning similarity data into discrete groups \citep{macqueen1967}. The optimal number of clusters was determined using the silhouette score, which evaluates the compactness and separation of the clusters \citep{rousseeuw1987}. The analysis revealed that a three-cluster solution provided a well-structured partitioning, as indicated by the silhouette score.

 We employed Non-negative Matrix Factorization (NMF) to extract representative temporal components associated with each cluster \citep{lee2000algorithms}. NMF is a dimensionality reduction technique that factorizes a non-negative matrix $X$ into two non-negative matrices, $W$ and $H$, such that:
\[
X \approx WH,
\]
where $X \in \mathbb{R}_{\geq 0}^{n \times t}$ is the matrix of node centrality time series, $W \in \mathbb{R}_{\geq 0}^{n \times r}$ contains the component weights for each node, and $H \in \mathbb{R}_{\geq 0}^{r \times t}$ contains the temporal profiles of each component. Where in our case $r$ is the number of diseases presenting at least one edge in across the seven age group, $t$ the number of age groups and $n$ the number of components identified by the k-means clustering (3).

We then plotted $H_i(t)$ for each component $i \in \{1, 2, 3\}$ to visualize how the associated patterns evolved over time. To enable direct comparison between components, we normalized each trajectory by its maximum value, i.e.,
\[
\widetilde{H}_i(t) = \frac{H_i(t)}{\max_t H_i(t)}.
\]

This normalization emphasizes the relative dynamics of each component across the time span under consideration.

\subsection*{Prevalence-Degree Correlation Analysis}

To assess how disease prevalence may influence the behavior of network properties, we explore the relationship between node degree and prevalence over time by analyzing its evolution across age groups \citep{divoCOPDComorbiditiesNetwork2015}. Let $d_i$ denote the degree of node $i$ in the network, and $p_i$ the prevalence associated with node $i$. We compute the log-ratio as:

\[
\text{Log-Ratio}_i = \log\left( \frac{d_i}{p_i} \right)
\]

To track potential changes, we focus on nodes that fall in the distribution's tails -- specifically, those below the 5th percentile and above the 95th percentile within its age group and sex. These extreme cases were examined further to identify potential outliers or structural shifts in the network.

\subsection*{Identification of High-Risk Diseases and Connections}

We analyzed the nodes' betweenness centrality in-hospital mortality across age groups. Our focus was on diagnoses exhibiting both high betweenness centrality and elevated in-hospital mortality rates. Specifically, we investigated how the position of these "high-in-hospital mortality bridges" changes over time, aiming to understand their shifting roles within the network structure.

To account for both high centrality and high mortality of nodes, we computed the product of the Z-Scores of these two features of nodes. The Z-Score of a value $x_i$ is defined as: $ z_i(x_i) = \frac{x_i - \mu}{\sigma}$, \( \mu \) is the mean of the distribution, \( \sigma \) is the standard deviation of the distribution. This score gives a standardized measure of how much an observed value deviates from the mean with respect to the standard deviation of the distribution.
Given the betweeness $g_i$ and the mortality $m_i$ of a node $i$, we computed $z(g_i)z(m_i)$ for all the diseases, and then selected the top $40\%$ of the distribution, which represents the nodes with both high centrality and mortality.

In \hyperref[sec:Results]{Results}, we report the geometric mean of these two scores, i.e. $\sqrt{z(g_i)z(m_i)}$, as we only focus on nodes with positive Z-Scores.




\subsection*{Robustness Analysis Across Co-Occurrence Metrics and Network Definitions} To validate the robustness of our results to the weights used to define comorbidities, we evaluate network properties over time using different co-occurrence metrics based on the paper \citep{monchka2022effect} -- \textbf{Lift}, \textbf{Relative Risk}, \textbf{Phi} ($\phi$), \textbf{Jaccard}, \textbf{Cosine}, \textbf{Kulczynski}, and \textbf{Joint Prevalence}. In this case, we compare across all networks: Female, Male, the outcome of such analysis is reported in \hyperref[sec:Supplement]{Supplementary Information}.

We also compute higher-order interactions measuring synergy and redundancy through O-information from the data of 14 chapters (see \hyperref[sec:Supplement]{Supplement Information}) \citep{rosas2019quantifying, neri2024hoi} .

\section{Results}\label{sec:Results}

We used a recently published dataset to construct comorbidity networks among 1,080 diseases across seven distinct age groups~\citep{dervic_comorbidity_2025}. In these networks, nodes represent diseases and edges represent statistically significant co-occurrence relationships between diseases, quantified using the odds ratio (see \hyperref[sec:Method]{Methods}). Previous studies have shown that the global topology of comorbidity networks undergoes abrupt transitions across different age groups~\citep{chmiel_spreading_2014}. To investigate how individual diseases contribute to this temporal evolution, we analyzed strength centrality, a measure for total edge weight, for each disease across different age groups.

\subsection*{Evolution of Node Strength and Centrality in Time}
\begin{figure}[!htbp]
    \centering
    \includegraphics[width=\linewidth]{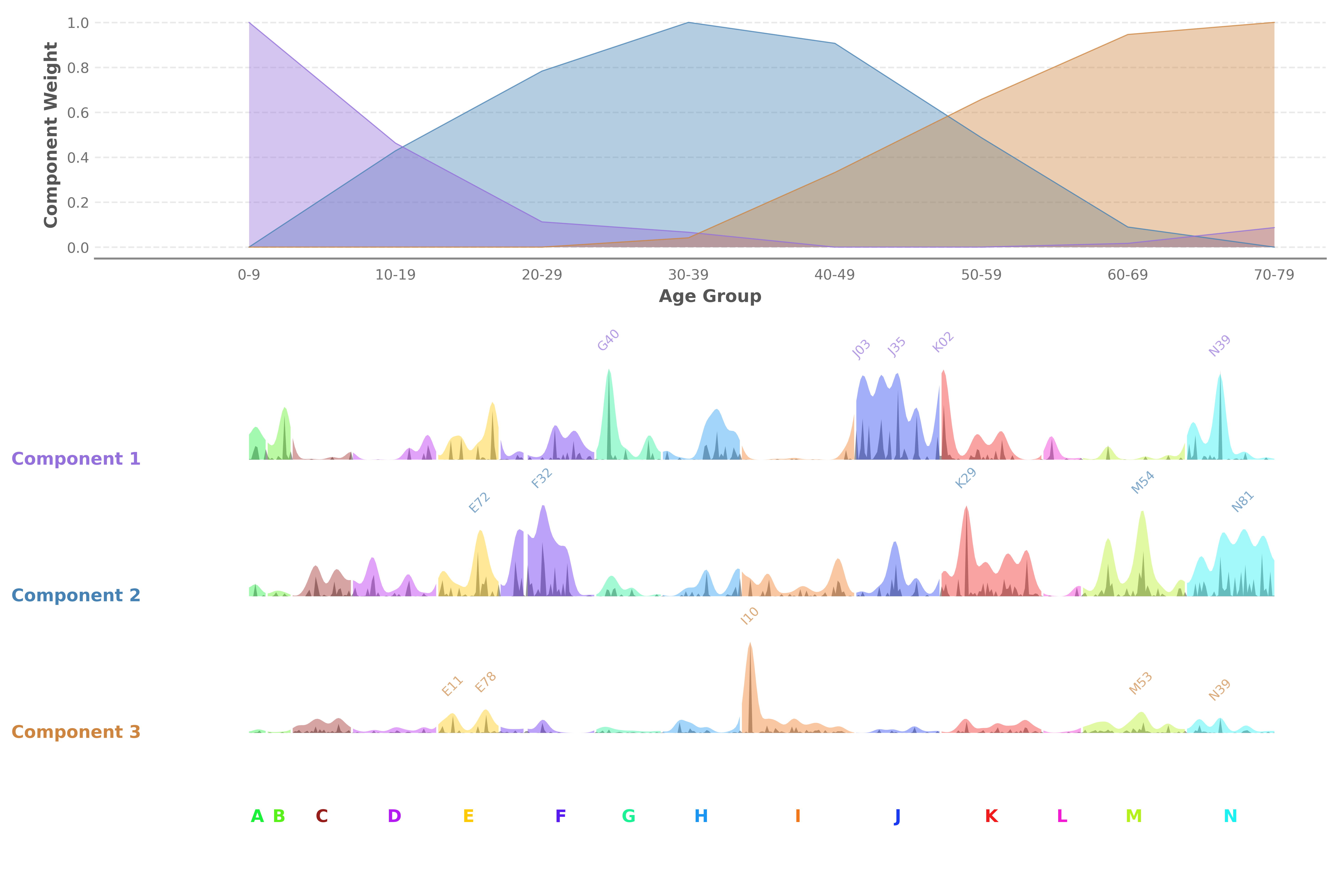}
\caption{\textbf{Temporal evolution of co-evolving disease clusters based on node strength centrality in comorbidity networks in females.} The top panel displays the normalized temporal profiles $\widetilde{H}_i(t)$ of three components extracted via Non-negative Matrix Factorization (NMF). The bottom panels show the contribution of individual diagnoses (ICD-10 codes) to each component across time, colored by ICD-10 chapter: A–B – Infectious and Parasitic Diseases, C – Neoplasms, D – Blood Diseases and Immune Disorders; E – Endocrine, Nutritional and Metabolic Diseases; F – Mental and Behavioural Disorders; G – Nervous System Diseases; H – Eye, Ear and Related Diseases; I – Circulatory System Diseases; J – Respiratory System Diseases; K – Digestive System Diseases; L – Skin and Subcutaneous Tissue Diseases; M – Musculoskeletal and Connective Tissue Diseases; N – Genitourinary System Diseases. Plot for males is  reported in the\hyperref[sec:Supplement]{Supplementary Information}.}
    \label{fig:temporal_evolution_female}
\end{figure}

We observed that distinct subsets of diseases exhibit similar temporal patterns of centrality. To systematically identify groups of diseases with similar strength centrality trajectories over time, we applied $k$-means clustering (see \hyperref[sec:Method]{Methods}), which revealed three major components of disease progression trends (Figure \ref{fig:female_kmeans_rational}). We then used non-negative matrix factorization (NMF) to decompose the strength centrality matrix and extract latent temporal patterns that contribute to the organization of the network. The top panel of Figure \ref{fig:temporal_evolution_female} displays the normalized temporal profiles $\widetilde{H}_i(t)$ of the three NMF components, each representing the prominence of a latent pattern across age groups. The bottom panels show the contribution of individual ICD-10 diagnoses to each NMF component. The diagnoses are colored by ICD-10 chapter and labeled with representative diagnoses among the top values of each component (see \hyperref[sec:Method]{Methods}).

More into details (see Figure \ref{fig:temporal_evolution_female}), the first component peaks at the age group 0-9 (purple), component 2 at the age group 30-39 (blue), and component 3 peaking in the age group 70-79 (orange). We find distinct patterns of diagnoses across the components. The 5 most predominant diagnoses in component 1 are epilepsy (G40), tonsillitis acute and chronic (J03 / J35), dental caries (K02), and urinary tract infection (N39) for females. 

For component 2 we find disorders of amino‑acid metabolism (E72), major depressive disorders (F32), gastritis and duodenitis (K29), low back pain (M54), and female genital prolapse (N81). 

For component 3 we find type 2 diabetes (E11), disorders of lipoprotein metabolism (E78), hypertension (I10), dorsopathies (M53), and and urinary tract infection (N39).

\subsection*{Prevalence-Degree Correlation Analysis}

Next, we asked which underlying factors might be causing the observed changes in disease patterns among different age groups. One plausible structural driver is the relationship between disease prevalence and degree, or how interconnected a disease is in the network. To determine whether increases in disease connectedness are merely a result of rising prevalence or if certain diseases become disproportionately connected over time, we examined the relationship between prevalence and degree across age groups.

There is a clear positive relationship between prevalence and degree across all age groups. More prevalent diseases tend to co-occur with more diseases. This reflects a fundamental property of comorbidity networks. A disease is more likely to be connected to others simply by appearing more often, although the weight of the edges partially compensates for this effect by representing the odds ratio between two diseases.

However, this relationship is not perfect. Some diseases deviate substantially from this trend. These deviations are meaningful because they point to conditions that are more or less connected than expected based on prevalence alone.

In younger age groups, most diseases are prevalent and have a low degree. The network is sparse, and the clusters remain relatively separate. As people age, the prevalence and degree of diseases increase, as does the number of high-degree nodes. The distribution of diseases becomes denser. This suggests that comorbidity networks become more entangled with age. Outliers with a high degree despite moderate prevalence may play an important role in this process.

\begin{figure}[!htbp]
    \centering
    \includegraphics[width=\linewidth]{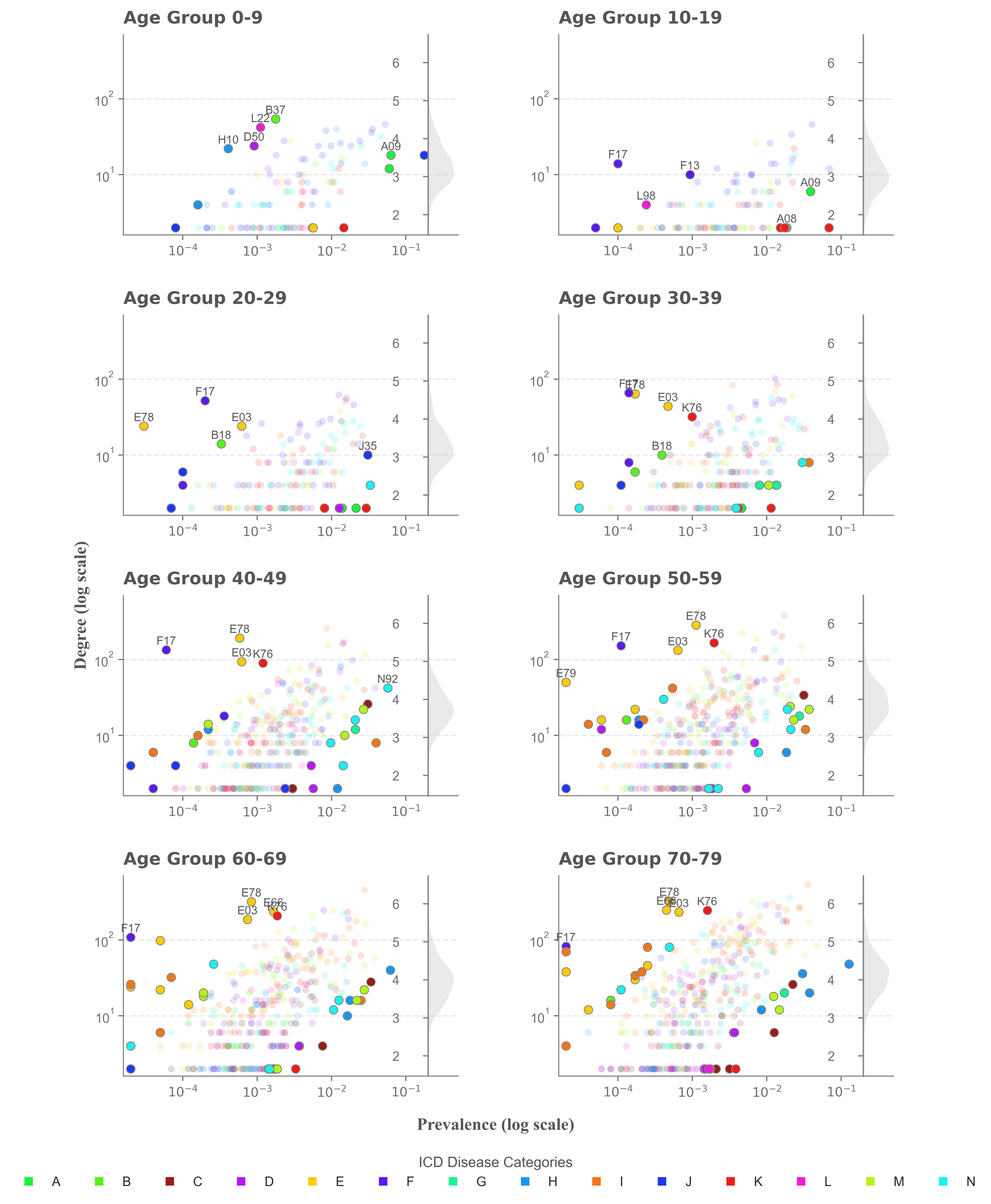}
\caption{\textbf{Outlier analysis across age groups for female patients.} Each subfigure shows a scatter plot of degree versus prevalence (on a log scale) for diseases in a given age group. Outliers are highlighted with solid markers, and the most significant ones have labels with their corresponding ICD codes. The right axis shows the distribution of log-transformed degree-to-prevalence ratios. Outliers vary significantly across age groups, reflecting how certain diseases' prevalence and network connectivity change depending on patient age. The plot for males is reported in the Supplementary Information.}
    \label{fig:prevalence_degree_female}
\end{figure}

To track structural shifts with age, we examine nodes that fall within the tails of the prevalence-degree distribution. Specifically, we focus on diagnoses below the 5th percentile or above the 95th percentile in each age group and sex. Diseases such as nicotine dependence (F17) and lipoprotein metabolism disorders (E78) consistently emerge as outliers in middle-aged and older adults. In children, we identified iron deficiency anemia (D50) as a notable outlier, showing a relatively high degree compared to its prevalence. These diagnoses exhibit a disproportionately high degree relative to their prevalence, suggesting strong correlations with a wide range of comorbid conditions. Notably, nicotine dependence becomes significantly more connected in older adults than in younger age groups, indicating an age-related increase in its comorbidity burden. Such diseases may play key structural roles in bridging otherwise distinct disease clusters and could serve as early indicators of emerging multimorbidity.

Conversely, other conditions such as eye-related diseases (H26, H35) show the opposite pattern—they are common among the elderly but have relatively low degree, indicating limited connectivity to other conditions. Similar outliers include infectious gastroenteritis and colitis (A09) in individuals under 20, and sleep disorders (G47) in adults—both of which exhibit high prevalence but low structural integration within the comorbidity network.

\subsection*{Detection of High-Mortality Sinks and Low-to-High Mortality Bridges}
\begin{figure}[!htbp]
    \centering
    \includegraphics[width=\linewidth]{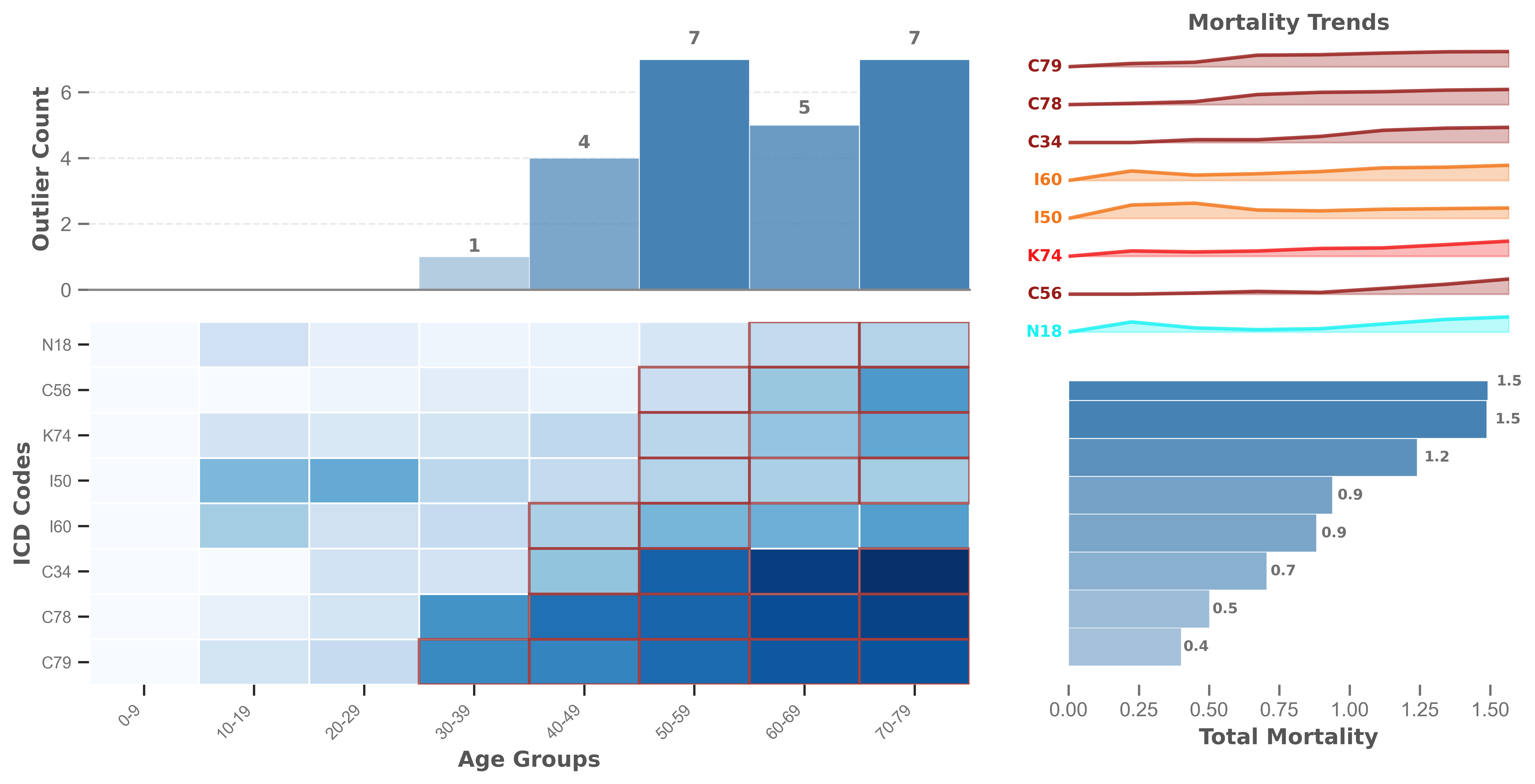}
\caption{\textbf{Evolution of high-mortality sinks diseases in females.} The top left panel shows the number of diagnoses with high mortality and betweenness -- identified by elevated geometric means of their mortality and betweenness Z-scores -- per age group. The bottom left heatmap highlights selected high-mortality sinks -- named according to their ICD-10 codes -- across age groups, with color intensity representing mortality rates and red borders denoting high betweenness too. The top right panel presents mortality trends over time for these diagnoses. The bottom right panel displays the total mortality score for each diagnosis, aggregated across all age groups.}
    \label{fig:high-mortality-bridges-females}
\end{figure}

The network structure analysis also allows us to identify diseases that constitutes "high-mortality sinks", as well as links that serve as "low-to-high mortality bridges".
We use betweeness centrality as a proxy to identify these high-mortality sinks. If a node has high betweeness, it means it lies on many of the shortest paths between other nodes, and is indeed central in the network structure.
If, additionally, that node also has a high mortality rate, we identify it as a high-mortality sink.
These nodes are particularly harmful, as there are many paths leading to them and they have high mortality.
In Figure \ref{fig:high-mortality-bridges-females}, we report the geometric mean of the Z-Scores of betweeness and mortality for the high-risk sink diseases as a function of age groups (see \hyperref[sec:Method]{Method} for a detailed description of how to identify them).
Not surprisingly, the highest risk sinks mostly belong to the cancer group of diseases. Not only these nodes are central in the network, and thus connected to many groups of diseases, but they also have high mortality, especially in older age groups (see Figure \ref{fig:high-mortality-bridges-females}).
Other important high-risk sinks are the ICD codes K74, I60, N18, namely fibrosis and cirrhosis of liver, nontraumatic subarachnoid hemorrhage, and chronic kidney disease.

As for the high-mortality bridges, we identify them as edges with a high betweenness centrality and a high difference in mortality between the nodes they connect, using again the product of their z-scores. Similarly to the previous case, these edges are very central in the network, and the increase in mortality at their ends constitutes a risk factor.
For example, we find that C34 (malignant neoplasm of bronchus and lung) and F17 (nicotine dependence) are high-risk bridges in age group 5 for both sexes, because F17 has a low mortality and C34 a high one, and, additionally, this edge position is very central in the network.
Similarly, males in age group 4, the edge between E78 (disorders of lipoprotein metabolism and other lipidemias) and F10 (alcohol related disorders) constitutes a high-mortality bridge.
Refer to Supplementary Information for a detailed description of high-risk bridges.

We find that various forms of malignant neoplasms (C79, C78, C34) start arising as high-risk bridging diseases (Figure  \ref{fig:high-mortality-bridges-females}) between 30 and 50 years old. Between 40 and 49, nontraumatic subarachnoid hemorrhage (I60), a cerebrovascular disease emerges as high-risk bridging diseases but disappear again starting from 60 to 69. At the age of 50 to 59 multiple disorders such as heart failure (I50), fibrosis and cirrhosis of the liver (K74), and ovarian cancer  (C56) emerge. Between 60 and 69, additionally chronic kidney disease (N18) emerges as a bridging disease.

Finally, we compare the outliers shown in Figures \ref{fig:prevalence_degree_female} and \ref{fig:prevalence_degree_male} to the high-risk diseases shown in Figure \ref{fig:high-mortality-bridges-females} to understand whether any diseases are both high-risk and also differ substantially from traditional comorbidity patterns.
The overlap between these classifications is small, but these few diseases are shown in Table \ref{tab:prevalence-outlier-risk-overlap}.

\begin{table}[!ht]
    \centering
    \begin{tabular}{lcclc}
        \toprule
        \textbf{Sex} & \textbf{Age} & \textbf{Jaccard} & \textbf{Intersection} & \textbf{Num. Outliers} \\
        \midrule
        Female & 1 & 0.14 & J90, D50, E10 & 12 \\
        Female & 2 & 0.00 & \textcolor{gray}{None} & 11 \\
        Female & 3 & 0.03 & E03 & 16 \\
        Female & 4 & 0.02 & E03 & 21 \\
        Female & 5 & 0.05 & E03, K76, C50 & 31 \\
        Female & 6 & 0.06 & E03, K76, C50, D48 & 35 \\
        Female & 7 & 0.06 & E03, E66, K76, C50 & 38 \\
        Female & 8 & 0.00 & \textcolor{gray}{None} & 40 \\
        Male & 1 & 0.13 & D50, E10, H10 & 13 \\
        Male & 2 & 0.00 & \textcolor{gray}{None} & 10 \\
        Male & 3 & 0.00 & \textcolor{gray}{None} & 12 \\
        Male & 4 & 0.04 & K76 & 14 \\
        Male & 5 & 0.04 & J43, K76 & 25 \\
        Male & 6 & 0.07 & B35, E79, J43, K76 & 31 \\
        Male & 7 & 0.09 & H25, C67, K76, H10, E79, C44 & 37 \\
        Male & 8 & 0.12 & H25, C67, H35, B35, K76, H10, E79, C44 & 37 \\
        \bottomrule
    \end{tabular}
    \caption{Comparing prevalence outliers and high-risk diseases. To make a comparison across ages, we consider the top $k$ diseases according to risk, where $k$ is the number of outliers identified in Figures \ref{fig:prevalence_degree_female} and \ref{fig:high-mortality-bridges-females} and \ref{fig:prevalence_degree_male}.}
    \label{tab:prevalence-outlier-risk-overlap}
\end{table}

\section{Discussion}

Network medicine has emerged in recent years as a powerful framework to study diseases, improve healthcare, and inform prevention strategies \citealp{cvijovic2023network, sonawane2019network, chan2012emerging, barabasi2011network}. In particular, comorbidity networks offer valuable insights into patterns of disease progression across a patient’s lifespan. For the first time, we utilized a public comorbidity network dataset released by \citealp{dervic_comorbidity_2025} to explore how network theory can reveal meaningful correlations between diseases and their relationships with age and sex.

At the macro level, our analysis showed that comorbidity networks become progressively larger and denser with age. From childhood (ages 0–9) to old age (ages 70–79), the female network grew from 111 nodes with an average degree of 5.00 to 384 nodes with an average degree of 21.78. A similar pattern emerged for males, whose network grew from 136 nodes with a degree of 6.25 in early childhood to 359 nodes with a degree of 20.39 in later life. Children tend to receive various diagnoses that remain structurally isolated, as reflected by higher modularity (0.49 for females and 0.43 for males) and longer average path lengths (3.56 for females and 3.24 for males) between the ages of 0 and 9. In contrast, older adults exhibited more entangled networks with higher connectivity and reduced modularity (0.24 for females and 0.25 for males aged 70–79). This suggests that aging is not merely the accumulation of diseases, but rather, it is defined by the growing interconnectedness of health conditions.

To better understand how these interconnections evolve over time, we used nonnegative matrix factorization (NMF) to extract dominant temporal patterns in disease centrality. This analysis revealed three distinct components, each of which characterized a life phase: childhood, midlife, and old age. This highlights how the structure of comorbidity changes with age and points to the emergence of age-specific disease clusters.

The three temporal components reflect distinct disease profiles emerging at different times in life, suggesting that diagnoses play a dynamic role in the comorbidity network. Some conditions like urinary tract infections (N39) appeared in childhood and old age components, while others like hypertension (I10) and metabolic disorders (E11, E78) emerged only in older age. These evolving roles raises a key question: are these diseases defining the components because they are more prevalent or do they influence the network structure regardless of frequency? 

To address this question, we examined the relationship between prevalence and degree across age groups. Our findings reveal that certain diseases are disproportionately connected within the comorbidity network and may play central roles in disease progression. For example, iron deficiency anemia (D50) has unusually high connectivity in the 0–9 age group, and lipoprotein metabolism disorders (E78) is a consistently highly connected node from early adulthood onward. Among adult females, hypothyroidism (E03) shows similarly elevated connectivity relative to its prevalence. These conditions may function as key hubs or multipliers in disease propagation and should therefore be prioritized in clinical risk stratification and healthcare planning.

In contrast, some prevalent conditions, such as infectious gastroenteritis and colitis (A09) in youth, sleep disorders (G47) in adults, and age-related eye diseases (H26/H35) in the elderly, exhibited a low degree of multimorbidity despite their frequent occurrence. These diagnoses may represent structurally isolated or self-limiting conditions that contribute less to broader multimorbidity patterns. Identifying these disconnects helps distinguish clinically significant diseases from prevalent but peripheral ones.


To better understand how disease interactions influence risk, we examined the structural role of diagnoses within the comorbidity network. We focused on diseases and connections with high betweenness centrality. By linking this measure to in-hospital mortality, we identified conditions that constitute high-mortality sinks. Similarly, we found high-mortality bridges, i.e. critical connectors on the path to severe health outcomes.
Cancers such as secondary malignant neoplasm (C78), secondary malignant neoplasm of other sites (C79), and lung cancer (C34) consistently demonstrate high structural embedding from midlife. These diagnoses also exhibited steep increases in mortality with age, marking them as high-mortality sinks. Other notable cases included subarachnoid hemorrhage (I60), ovarian cancer (C56), and chronic kidney disease (N18), which increased in both structural influence and mortality. Together, these patterns highlighted diseases that are central in the network and associated with severe outcomes, underscoring their importance as targets for prevention or early intervention.

The limitations of our study mostly concern the features of the dataset.
The data are hospital-based and lack longitudinal tracking at the patient level. Additionally, disease coding is limited to ICD categories, which may obscure finer clinical distinctions. Additionally, observed comorbidities may arise from shared risk factors, treatment effects, or documentation practices rather than true biological associations. Additionally, since the data were originally collected for administrative and billing purposes, certain diagnoses may be over- or underrepresented. Another limitation is that mortality is only captured as in-hospital mortality. This excludes deaths occurring outside the hospital and may lead to an underestimation of the true mortality burden associated with specific conditions. From a methodological standpoint, using odds ratios to define network edges does not imply causality. For example, a strong link between dyslipidemia (E78) and primary hypertension (I10) does not clarify whether one causes the other or if they simply co-occur frequently.

Future research could address these limitations by integrating additional data sources such as incorporating medication and electronic health records with longitudinal follow-up and established causal relationships between diseases. These additions could support the construction of directed or signed networks, enabling the representation of causal or protective relationships between conditions. This would be a crucial step toward creating actionable, predictive comorbidity models.

Ultimately, our findings reinforce the value of network medicine in capturing the evolving complexity of disease interactions across the human lifespan. Age-specific comorbidity networks highlight not only how diseases co-occur but also how these interactions evolve with age. This dynamic perspective could inform more personalized and proactive healthcare strategies.






 


\section{Acknowledgements}
This work is the output of the Complexity72h workshop, held at the Universidad Carlos III de Madrid in Leganés, Spain, 23-27 June 2025. https://www.complexity72h.com

\section{Supplementary Information} \label{sec:Supplement}
\subsection*{Overview of the ICD-10 Chapters}
\begin{table}[h!]
\centering
\begin{tabular}{cl}
\toprule
\textbf{ICD-10 Letter(s)} & \textbf{Chapter Name} \\
\midrule
A--B     & Infectious \& Parasitic Diseases \\
C        & Neoplasms \\
D        & Blood Diseases \& Immune Disorders \\
E        & Endocrine, Nutritional \& Metabolic Diseases \\
F        & Mental \& Behavioural Disorders \\
G        & Nervous System Diseases \\
H        & Eye, Ear \& Related Diseases \\
I        & Circulatory System Diseases \\
J        & Respiratory System Diseases \\
K        & Digestive System Diseases \\
L        & Skin \& Subcutaneous Tissue Diseases \\
M        & Musculoskeletal \& Connective Tissue Diseases \\
N        & Genitourinary System Diseases \\
\bottomrule
\end{tabular}
\caption{ICD-10 Chapters A--N}
\end{table}

\renewcommand{\thefigure}{S\arabic{figure}}
\setcounter{figure}{0}
\subsection*{Reconstruction error and evaluation of clustering}

\textbf{Silhouette Analysis}

To evaluate the quality of clustering, we computed the \emph{silhouette score}, which quantifies the consistency of data point assignments within clusters. For each data point \( i \), the silhouette score \( s(i) \) is defined as:

\begin{equation}
s(i) = \frac{b(i) - a(i)}{\max \{a(i), b(i)\}}
\end{equation}

where:
\begin{itemize}
    \item \( a(i) \) is the mean intra-cluster distance: the average distance between \( i \) and all other points in the same cluster.
    \item \( b(i) \) is the mean nearest-cluster distance: the minimum average distance between \( i \) and all points in the nearest neighboring cluster (i.e., the closest cluster to which \( i \) does \emph{not} belong).
\end{itemize}

The silhouette score ranges from \(-1\) to \(+1\), with higher values indicating better-defined clusters. A value close to \( +1 \) implies that the point is well matched to its own cluster and poorly matched to others; values around \( 0 \) indicate ambiguity near cluster boundaries, while values near \( -1 \) suggest misclassification.

The overall clustering performance was assessed by computing the mean silhouette score across all data points. All silhouette scores were calculated using the \texttt{silhouette\_score} function from the \texttt{scikit-learn} library.

\begin{figure}[!htbp]
    \centering
\includegraphics[width=\linewidth] {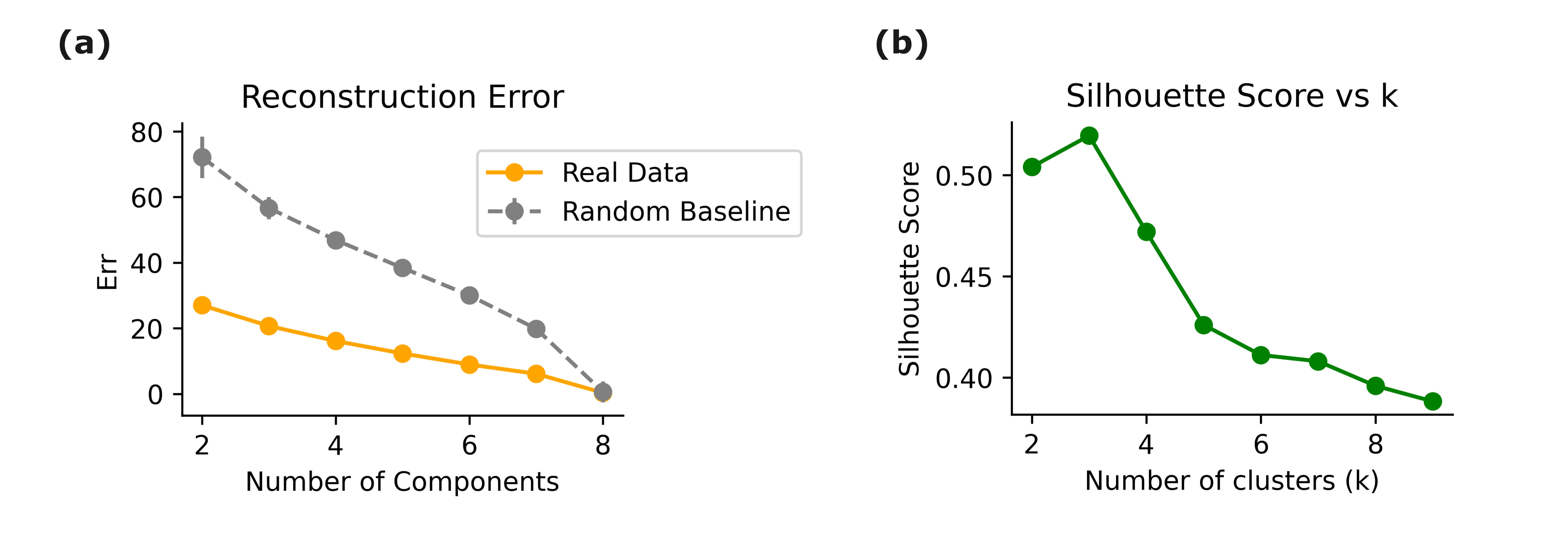}
\caption{\textbf{Reconstruction error and evaluation of clustering.} Here, we report the reconstruction error in comparison to a null model obtained by randomly shuffling the data (Panel a). In Panel b, we present the silhouette plot to assess the quality of the clustering solution.}
    \label{fig:female_kmeans_rational}
\end{figure}

\begin{figure}[!htbp]
    \centering
\includegraphics[width=0.8\linewidth] {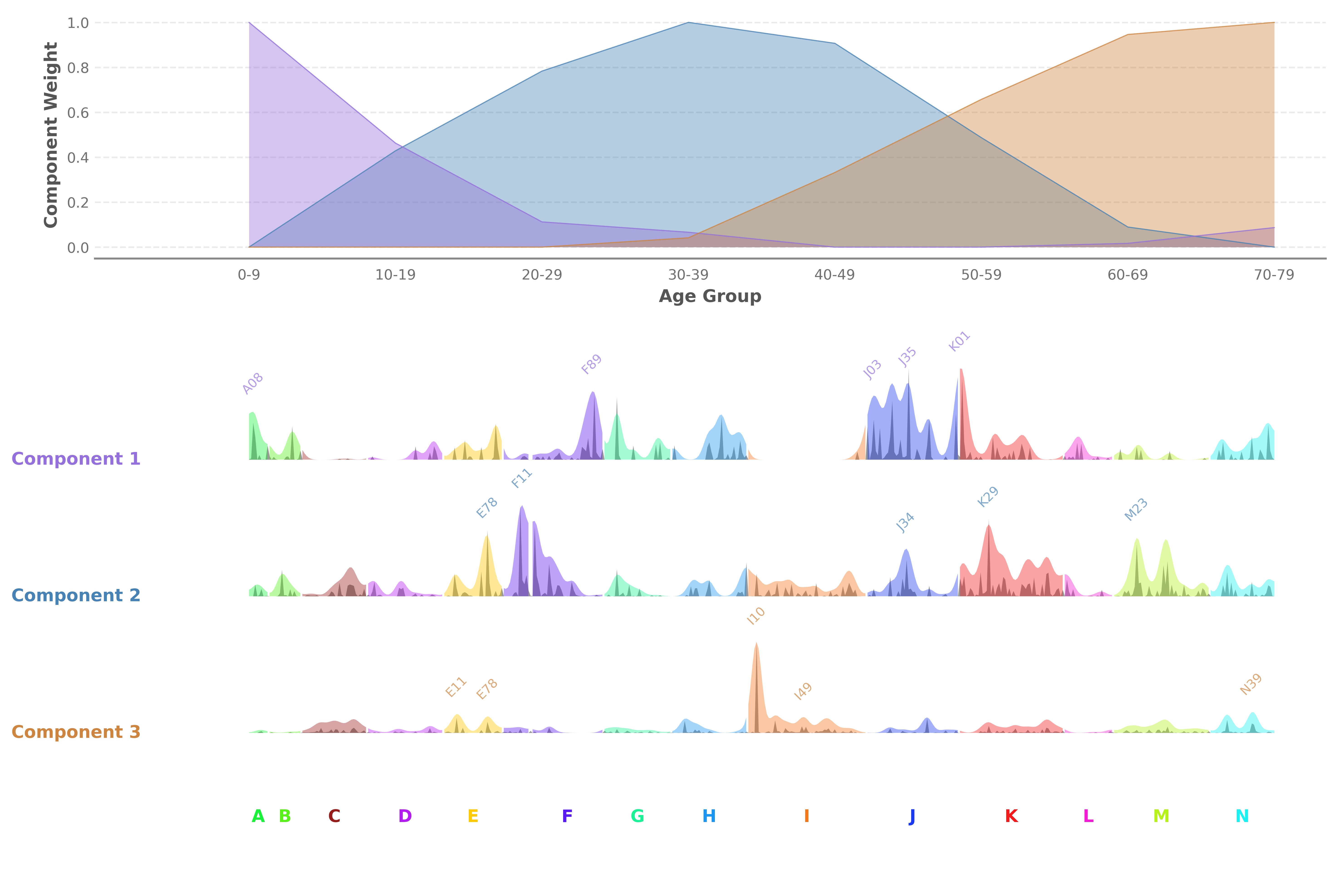}
\caption{\textbf{Temporal evolution of co-evolving disease clusters based on node strength centrality in comorbidity networks in males.} The top panel displays the normalized temporal profiles $\widetilde{H}_i(t)$ of three components extracted via Non-negative Matrix Factorization (NMF). The bottom panels show the contribution of individual diagnoses (ICD-10 codes) to each component across time, colored by ICD chapter.}
    \label{fig:temporal_evolution_male}
\end{figure}

\newpage
\subsection*{Role of Disease Position in Comorbidity Networks}

\begin{figure}[!htbp]
    \centering
    \includegraphics[width=0.9\linewidth]{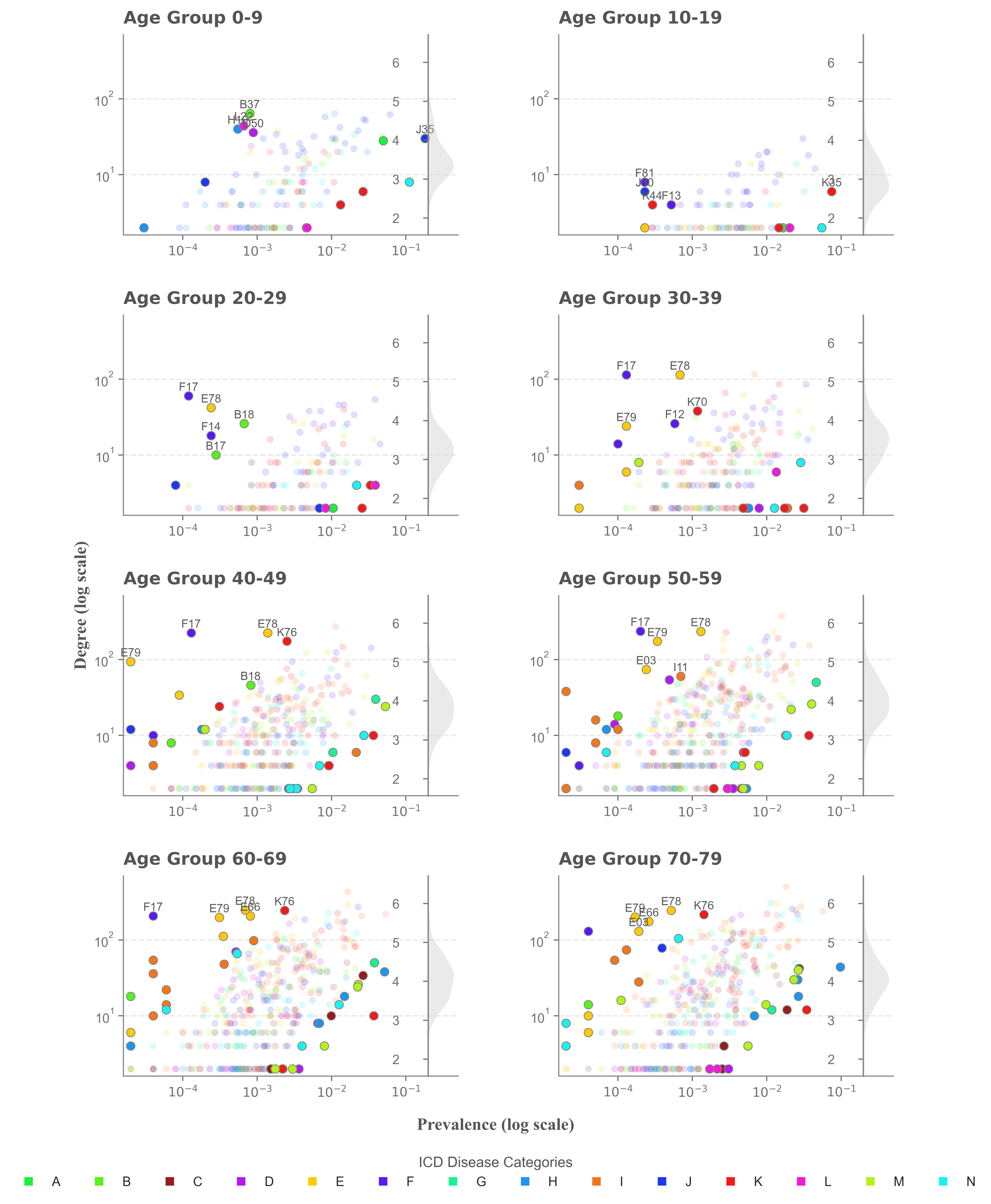}
\caption{\textbf{Outlier analysis across age groups for male patients.}}
    \label{fig:prevalence_degree_male}
\end{figure}

\newpage

\subsection*{Metric-Based Edge Weighting of Comorbidity Networks}
The formulas for each metric used to validate the robustness of our results to the weights used to define comorbidities are as follows:

\begin{itemize}
    \item \textbf{Lift}:
    \[
    \text{Lift} = n \cdot \frac{a}{(a + b)(a + c)}
    \]

    \item \textbf{Relative Risk}:
    \[
    \text{RR} = \frac{a(c + d)}{c(a + b)}
    \]

    \item \textbf{Phi coefficient} ($\phi$):
    \[
    \phi = \frac{ad - bc}{\sqrt{(a + b)(a + c)(b + d)(c + d)}}
    \]

    \item \textbf{Jaccard Index}:
    \[
    J = \frac{a}{a + b + c}
    \]

    \item \textbf{Cosine Similarity}:
    \[
    \text{Cosine} = \frac{a}{\sqrt{(a + b)(a + c)}}
    \]

    \item \textbf{Kulczynski Index}:
    \[
    K = \frac{1}{2} \left( \frac{a}{a + b} + \frac{a}{a + c} \right)
    \]

    \item \textbf{Joint Prevalence}:
    \[
    JP = \frac{a}{n}
    \]
\end{itemize}
\noindent The metrics are based on the 2x2 contingency table structure:

\begin{center}
\begin{tabular}{@{}lcc@{}}
\toprule
 & \textbf{Diagnosis B Present} & \textbf{Diagnosis B Absent} \\
\midrule
\textbf{Diagnosis A Present} & $a$ & $b$ \\
\textbf{Diagnosis A Absent}  & $c$ & $d$ \\
\bottomrule
\end{tabular}
\end{center}

\noindent where the parameters are defined as followed:
\begin{itemize}
    \item $a$: Number of individuals with both Diagnosis A and Diagnosis B
    \item $b$: Number of individuals with Diagnosis A but not Diagnosis B
    \item $c$: Number of individuals with Diagnosis B but not Diagnosis A
    \item $d$: Number of individuals with neither Diagnosis A nor Diagnosis B
    \item $n$: Total number of individuals ($n = a + b + c + d$)
\end{itemize}

\subsection*{Bridging Edges}
\begin{table}[ht]
\centering
\begin{tabular}{rlccclccc}
  \hline
 & \textbf{Sex} & \textbf{Age} & \textbf{Mortality\_i} & \textbf{Mortality\_j} & \textbf{Link} & \textbf{Product} & \textbf{Ebtw} & \textbf{Difference in mortality rates} \\ 
  \hline
1 & Female & 1.00 & 0.01 & 0.14 & G40\_J96 & 13.04 & 0.00 & 0.13 \\ 
  2 & Female & 1.00 & 0.00 & 0.17 & H66\_H70 & 7.19 & 0.00 & 0.17 \\ 
  3 & Female & 1.00 & 0.04 & 0.14 & J20\_J96 & 6.36 & 0.00 & 0.10 \\ 
  4 & Female & 2.00 & 0.06 & 0.01 & J06\_K59 & 1.35 & 0.00 & 0.05 \\ 
  5 & Female & 3.00 & 0.02 & 0.21 & E66\_I26 & 3.47 & 0.00 & 0.19 \\ 
  6 & Female & 3.00 & 0.07 & 0.00 & F50\_K29 & 1.21 & 0.00 & 0.07 \\ 
  7 & Female & 4.00 & 0.08 & 0.01 & C50\_F43 & 4.46 & 0.00 & 0.06 \\ 
  8 & Female & 4.00 & 0.00 & 0.05 & D25\_D64 & 2.57 & 0.00 & 0.05 \\ 
  9 & Female & 4.00 & 0.07 & 0.21 & F17\_I25 & 1.96 & 0.00 & 0.14 \\ 
  10 & Female & 4.00 & 0.04 & 0.00 & I10\_M54 & 0.62 & 0.00 & 0.04 \\ 
  11 & Female & 4.00 & 0.08 & 0.02 & C50\_D24 & 0.56 & 0.00 & 0.06 \\ 
  12 & Female & 4.00 & 0.00 & 0.00 & M24\_M65 & 0.54 & 0.00 & 0.00 \\ 
  13 & Female & 4.00 & 0.00 & 0.00 & K42\_K80 & 0.54 & 0.00 & 0.00 \\ 
  14 & Female & 5.00 & 0.35 & 0.07 & C34\_F17 & 7.87 & 0.00 & 0.28 \\ 
  15 & Female & 6.00 & 0.31 & 0.00 & C78\_K29 & 19.93 & 0.00 & 0.31 \\ 
  16 & Female & 7.00 & 0.31 & 0.02 & C78\_N39 & 27.41 & 0.00 & 0.30 \\ 
  17 & Female & 7.00 & 0.31 & 0.05 & C78\_D48 & 12.06 & 0.00 & 0.26 \\ 
  18 & Male & 1.00 & 0.19 & 0.49 & E86\_J96 & 2.75 & 0.00 & 0.30 \\ 
  19 & Male & 2.00 & 0.09 & 0.00 & F10\_F91 & 4.05 & 0.00 & 0.09 \\ 
  20 & Male & 3.00 & 0.09 & 0.00 & F10\_F41 & 3.09 & 0.00 & 0.09 \\ 
  21 & Male & 3.00 & 0.00 & 0.14 & J01\_J32 & 1.95 & 0.00 & 0.14 \\ 
  22 & Male & 3.00 & 0.09 & 0.02 & F10\_F17 & 0.56 & 0.00 & 0.07 \\ 
  23 & Male & 4.00 & 0.00 & 0.09 & E78\_F10 & 8.51 & 0.00 & 0.09 \\ 
  24 & Male & 4.00 & 0.09 & 0.01 & F10\_K02 & 3.91 & 0.00 & 0.08 \\ 
  25 & Male & 4.00 & 0.00 & 0.09 & E78\_G40 & 1.53 & 0.00 & 0.09 \\ 
  26 & Male & 4.00 & 0.14 & 0.01 & J32\_J45 & 1.43 & 0.00 & 0.13 \\ 
  27 & Male & 5.00 & 0.43 & 0.02 & C34\_F17 & 48.61 & 0.00 & 0.41 \\ 
  28 & Male & 6.00 & 0.39 & 0.02 & C78\_F17 & 25.08 & 0.00 & 0.37 \\ 
  29 & Male & 6.00 & 0.43 & 0.09 & C34\_G40 & 14.79 & 0.00 & 0.34 \\ 
  30 & Male & 7.00 & 0.43 & 0.02 & C34\_D48 & 29.64 & 0.00 & 0.41 \\ 
  31 & Male & 7.00 & 0.43 & 0.03 & C34\_F32 & 18.38 & 0.00 & 0.40 \\ 
  32 & Male & 8.00 & 0.39 & 0.00 & C78\_K21 & 12.81 & 0.00 & 0.39 \\ 
  33 & Male & 8.00 & 0.40 & 0.01 & C25\_E11 & 7.99 & 0.00 & 0.39 \\ 
   \hline
\end{tabular}
\end{table}

\subsection*{Sensitivity of Betweenness and Modularity to Edge Weighting Schemes}

The overall trends in betweenness centrality and modularity are consistent across the different weighting metrics, indicating that our results are robust to the choice of comorbidity measure. While the absolute values of network metrics vary slightly depending on the edge definition, the relative rankings of high-betweenness nodes remain stable. This suggests that the structural backbone of the comorbidity networks—particularly the identification of central bridging diseases is preserved across alternative co-occurrence metrics. 

\begin{figure}[t]
    \centering
    \includegraphics[width=\linewidth]{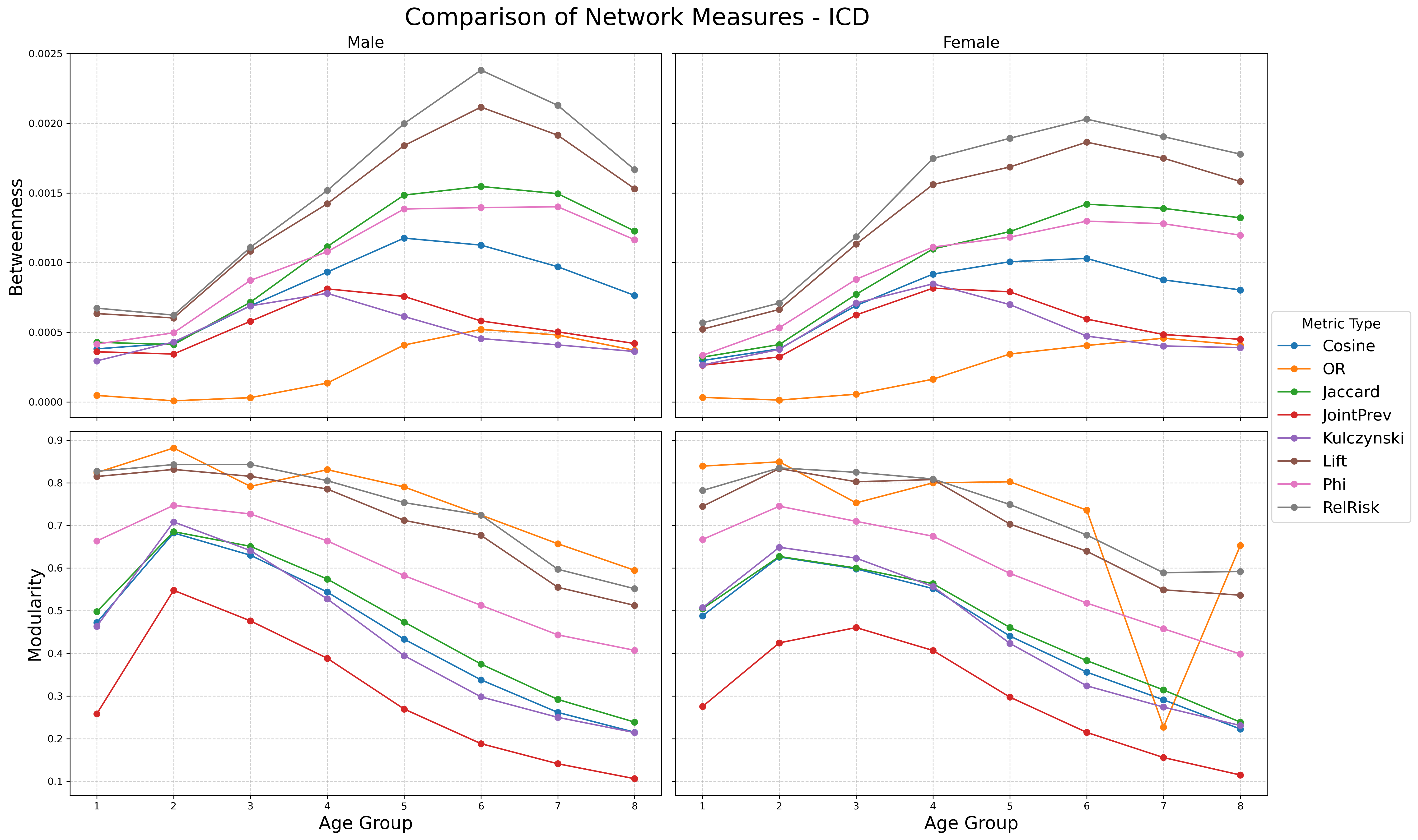}
\caption{\textbf{Comparison of network metrics betweenness and modularity across edge weighting schemes:} The plots show the values of \textbf{betweenness centrality} (top row) and \textbf{modularity} (bottom row) computed on the comorbidity networks constructed with different co-occurence metrics used as edge weights (indicated by line color). Each panel corresponds to a gender: \textbf{Male (left)}, and \textbf{Female (right)}. The x-axis represents age groups, allowing for the analysis of metric behavior across age. This visualization assesses the sensitivity of network structure to the choice of weight definition over time and by sex.}
    \label{fig:similarity_metrics}
\end{figure}

\subsection*{Higher-Order Analysis}
To gain a deeper understanding of the temporal dynamics underlying comorbility network changes, we extended our analysis beyond pairwise associations. These metrics provide a new lens through which to examine the evolving interplay between disease categories across the lifespan, complementing traditional network centrality measures~\citep{rosas2019quantifying}. Moreover, higher-order analysis shown to be relevant in different fields from neuroscience~\citep{neri2023neuronal, neri2025taxonomy, combrisson2024higher, coronel2025integrated, runfola2025complexity} to economy~\citep{scagliarini2023gradients}. Here, we employed recent advances in information theory to investigate higher-order interactions in terms of synergy and redundancy \citep{rosas2019quantifying}. For this analysis the diseases were grouped into 14 distinct categories, based on previous literature.

Interestingly, we observed distinct age-related patterns in these two informational components. Redundancy—reflecting overlapping, possibly substitutable contributions from co-occurring diseases—tended to peak during early and late life stages. In contrast, synergy—indicating unique, joint contributions beyond pairwise effects, displayed on average a more stable profile over time. This dual behavior suggests that disease co-occurrence in childhood and late adulthood may be driven by more stereotyped, redundant pathophysiological mechanisms (e.g., shared risk factors or diagnostic biases), whereas synergy may reflect more complex, system-level interactions emerging in adulthood~\citep{robiglio2025synergistic, marinazzo2019synergy}.

To further dissect these dynamics, we tracked the temporal evolution of synergy and redundancy at the level of individual nodes (i.e., diseases), grouped by their ICD-10 chapters \citep{Icd10}. Notably, several disease chapters that exhibited high centrality in the pairwise comorbidity network—such as \textit{A} (infectious and parasitic diseases), \textit{B} (blood disorders), \textit{J} (respiratory diseases), \textit{H} (eye and ear diseases), and \textit{M} (musculoskeletal disorders)—showed divergent behaviors in terms of their synergy/redundancy profiles. Specifically, the early-life centrality of chapters \textit{M} and \textit{H} appeared to be driven primarily by high redundancy, suggesting that comorbidities in these domains may reflect overlapping or substitutive mechanisms (e.g., multiple musculoskeletal symptoms linked to shared biomechanical stress or developmental factors). In contrast, the centrality of chapters \textit{B} and \textit{J} was more strongly associated with synergy, indicating that these conditions contribute unique, non-redundant information to disease interactions during early development—possibly due to their intersection with immune and respiratory system maturation.

Chapters \textit{A} and \textit{J} demonstrated a mixture of synergistic and redundant contributions, suggesting a more complex and young age-dependent modulation of their role in the comorbidity structure.

As individuals age, redundancy increases across nearly all disease chapters, but only in the two last age groups (60–79). This may reflect the cumulative burden of multimorbidity, where overlapping disease mechanisms—such as inflammation, metabolic dysregulation, or vascular compromise—converge and reinforce redundant patterns. However, synergy does not uniformly increase; instead, it peaks at intermediate life stages for a subset of disease chapters including \textit{F} (mental and behavioral disorders), \textit{L} (skin), \textit{D} (blood), \textit{K} (digestive), and \textit{M} (musculoskeletal), which are also highly central in the pairwise comorbidity network. This observation aligns with previous studies showing that \emph{synergistic information tends to increase during phase transitions} in complex systems, supporting the hypothesis that mid-life represents a transitional state in the human health trajectory—where interactions between systems acquire a different behavior \cite{marinazzo2019synergy, robiglio2025synergistic}. Importantly, many of these mid-life-synergistic chapters involve systems with strong biopsychosocial interactions (e.g., mental health, skin, digestive health), suggesting that synergy may reflect not only biological complexity but also psychosomatic and lifestyle-mediated effects.

In later life stages, we identified chapters \textit{C} (neoplasms), \textit{M}, and \textit{H} as maintaining a predominantly synergistic interaction profile, whereas other chapters became increasingly redundant. The persistent synergy of these chapters may relate to their involvement in multi-system conditions or treatment-driven effects (e.g., cancer and its systemic impact; sensory degradation influencing cognition and mobility; musculoskeletal disorders affecting physical and social function). Their pivotal role in late-life comorbidity structure may therefore stem from their capacity to act as integrative hubs within the broader health landscape, amplifying the complexity of interactions even as redundancy rises elsewhere.

\begin{figure}[!htbp]
    \centering
    \includegraphics[width=\linewidth]{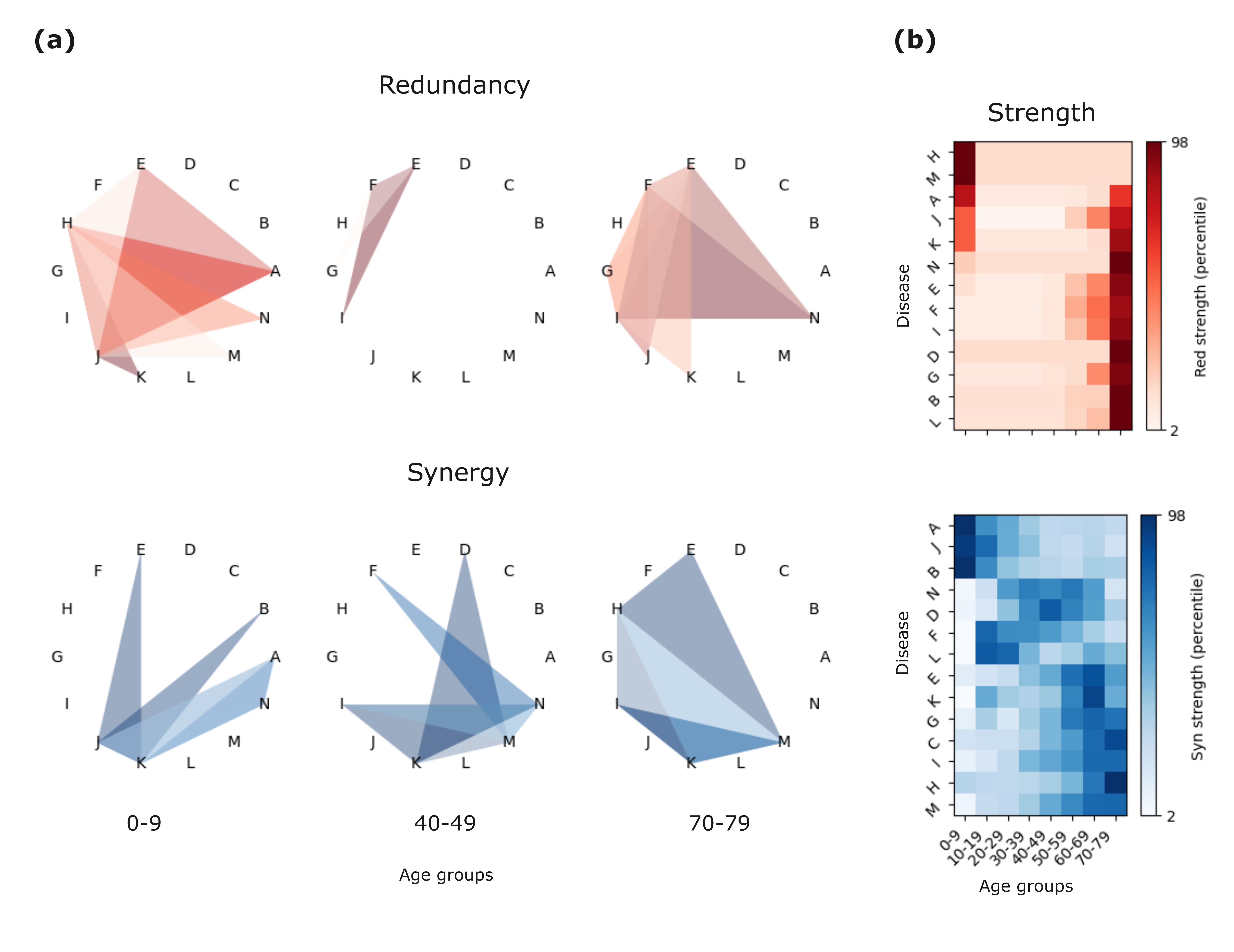}
\label{fig:higher-horder}\caption{\textbf{Higher-order dynamics of disease evolution.} To efficiently quantify higher-order interactions, diseases were grouped into 14 clinically relevant categories. We then computed synergy and redundancy using the O-information framework \citep{rosas2019quantifying} (see methods). Panel (a) displays the five strongest synergy- and redundancy-dominated hyperedges across three age groups, highlighting how higher-order dependencies vary over the lifespan. To further investigate the changing roles of individual disease groups, we projected the results to the node level, illustrating how each group's synergy and redundancy evolve with age, panel (b). }
\end{figure}

\subsection*{Higher-order analysis methods}

To characterize higher-order statistical dependencies among groups of diseases, we employed information-theoretic measures based on the O-information framework. Let \( \mathbf{X}_n = (X_1, \dots, X_n) \) denote a set of \( n \) random variables. The \emph{total correlation} (TC) and \emph{dual total correlation} (DTC) are defined respectively as:

\begin{align}
\mathrm{TC}(\mathbf{X}_n) &= \sum_{i=1}^n H(X_i) - H(\mathbf{X}_n) \\
\mathrm{DTC}(\mathbf{X}_n) &= H(\mathbf{X}_n) - \sum_{i=1}^n H(X_i \mid \mathbf{X}_{-i})
\end{align}

Here, \( H(\cdot) \) denotes the Shannon entropy, and \( \mathbf{X}_{-i} \) is the set of all variables in \( \mathbf{X}_n \) excluding \( X_i \). Both TC and DTC are non-negative and equal zero only when all variables are statistically independent.

The \emph{O-information}, \( \Omega(\mathbf{X}_n) \), is computed as the difference between TC and DTC:

\begin{equation}
\Omega(\mathbf{X}_n) = \mathrm{TC}(\mathbf{X}_n) - \mathrm{DTC}(\mathbf{X}_n)
\end{equation}

This quantity reflects the balance between redundancy and synergy among the variables~\citep{rosas2019quantifying}. A positive value (\( \Omega > 0 \)) indicates redundancy-dominated interactions (\emph{O-Red}), while a negative value (\( \Omega < 0 \)) indicates synergy-dominated interactions (\emph{O-Syn}).

In the special case of \( n = 3 \), the O-information reduces to the classical interaction information and can be expressed as:

\begin{align}
\Omega(X_i, X_j, X_k) &= H(X_i) + H(X_j) + H(X_k) \nonumber \\
&\quad - H(X_i, X_j) - H(X_i, X_k) - H(X_j, X_k) + H(X_i, X_j, X_k)
\end{align}

All information-theoretic computations were carried out using the \texttt{HOI} toolbox \cite{neri2024hoi}, specifically the \texttt{Oinfo} class of functions, available at: \url{https://github.com/brainets/hoi.io}. In our analysis we considered as redundant and synergistic tripets, respectively, the positive and negative values of O-information.


\begin{thebibliography}{53}
\providecommand{\natexlab}[1]{#1}
\providecommand{\url}[1]{\texttt{#1}}
\expandafter\ifx\csname urlstyle\endcsname\relax
  \providecommand{\doi}[1]{doi: #1}\else
  \providecommand{\doi}{doi: \begingroup \urlstyle{rm}\Url}\fi

\bibitem[Aguado et~al.(2020)Aguado, Moratalla-Navarro, López-Simarro, and
  Moreno]{aguado_morbinet_2020}
A.~Aguado, F.~Moratalla-Navarro, F.~López-Simarro, and V.~Moreno.
\newblock {MorbiNet}: multimorbidity networks in adult general population.
  {Analysis} of type 2 diabetes mellitus comorbidity.
\newblock \emph{Scientific Reports}, 10\penalty0 (1):\penalty0 2416, Feb. 2020.
\newblock ISSN 2045-2322.
\newblock \doi{10.1038/s41598-020-59336-1}.
\newblock URL \url{https://doi.org/10.1038/s41598-020-59336-1}.

\bibitem[Bao et~al.(2023)Bao, Lu, Wang, Zhang, Song, Gu, Ma, Su, Wang, Shang,
  Zhu, Zhai, He, Li, Liu, Fairley, Yang, and Zhang]{bao_exploring_2023}
Y.~Bao, P.~Lu, M.~Wang, X.~Zhang, A.~Song, X.~Gu, T.~Ma, S.~Su, L.~Wang,
  X.~Shang, Z.~Zhu, Y.~Zhai, M.~He, Z.~Li, H.~Liu, C.~K. Fairley, J.~Yang, and
  L.~Zhang.
\newblock Exploring multimorbidity profiles in middle-aged inpatients: a
  network-based comparative study of {China} and the {United} {Kingdom}.
\newblock \emph{BMC Medicine}, 21\penalty0 (1):\penalty0 495, Dec. 2023.
\newblock ISSN 1741-7015.
\newblock \doi{10.1186/s12916-023-03204-y}.
\newblock URL \url{https://doi.org/10.1186/s12916-023-03204-y}.

\bibitem[Barab{\'a}si et~al.(2011)Barab{\'a}si, Gulbahce, and
  Loscalzo]{barabasi2011network}
A.-L. Barab{\'a}si, N.~Gulbahce, and J.~Loscalzo.
\newblock Network medicine: a network-based approach to human disease.
\newblock \emph{Nature reviews genetics}, 12\penalty0 (1):\penalty0 56--68,
  2011.

\bibitem[Beard et~al.(2016)Beard, Officer, de~Carvalho, Sadana, Pot, Michel,
  Lloyd-Sherlock, Epping-Jordan, Peeters, Mahanani, Thiyagarajan, and
  Chatterji]{beardWorldReportAgeing2016}
J.~R. Beard, A.~Officer, I.~A. de~Carvalho, R.~Sadana, A.~M. Pot, J.-P. Michel,
  P.~Lloyd-Sherlock, J.~E. Epping-Jordan, G.~M. E. E.~G. Peeters, W.~R.
  Mahanani, J.~A. Thiyagarajan, and S.~Chatterji.
\newblock The {{World}} report on ageing and health: A policy framework for
  healthy ageing.
\newblock \emph{The Lancet}, 387\penalty0 (10033):\penalty0 2145--2154, 2016.
\newblock ISSN 0140-6736, 1474-547X.
\newblock \doi{10.1016/S0140-6736(15)00516-4}.
\newblock URL
  \url{https://www.thelancet.com/journals/lancet/article/PIIS0140-6736(15)00516-4/fulltext?rss%3Dyes=&code=lancet-site}.

\bibitem[Bramesh and KM(2023)]{mConstructingDiseaseComorbidity2023}
S.~Bramesh and A.~K. KM.
\newblock Constructing disease comorbidity networks using statistical methods
  from electronic health records.
\newblock In \emph{2023 {{International Conference}} on {{Network}},
  {{Multimedia}} and {{Information Technology}} ({{NMITCON}})}, pages 1--6,
  2023.
\newblock \doi{10.1109/NMITCON58196.2023.10276292}.
\newblock URL \url{https://ieeexplore.ieee.org/abstract/document/10276292}.

\bibitem[Capobianco and Liò(2015)]{capobiancoComorbidityNetworksDisease2015}
E.~Capobianco and P.~Liò.
\newblock Comorbidity networks: Beyond disease correlations.
\newblock \emph{Journal of Complex Networks}, 3\penalty0 (3):\penalty0
  319--332, 2015.
\newblock ISSN 2051-1329.
\newblock \doi{10.1093/comnet/cnu048}.
\newblock URL \url{https://ieeexplore.ieee.org/abstract/document/8155946}.

\bibitem[Chan and Loscalzo(2012)]{chan2012emerging}
S.~Y. Chan and J.~Loscalzo.
\newblock The emerging paradigm of network medicine in the study of human
  disease.
\newblock \emph{Circulation research}, 111\penalty0 (3):\penalty0 359--374,
  2012.

\bibitem[Chmiel et~al.(2014)Chmiel, Klimek, and Thurner]{chmiel_spreading_2014}
A.~Chmiel, P.~Klimek, and S.~Thurner.
\newblock Spreading of diseases through comorbidity networks across life and
  gender.
\newblock \emph{New Journal of Physics}, 16\penalty0 (11):\penalty0 115013,
  2014.
\newblock ISSN 1367-2630.
\newblock \doi{10.1088/1367-2630/16/11/115013}.
\newblock URL \url{https://dx.doi.org/10.1088/1367-2630/16/11/115013}.
\newblock Publisher: IOP Publishing.

\bibitem[Combrisson et~al.(2024)Combrisson, Basanisi, Neri, Auzias, Petri,
  Marinazzo, Panzeri, and Brovelli]{combrisson2024higher}
E.~Combrisson, R.~Basanisi, M.~Neri, G.~Auzias, G.~Petri, D.~Marinazzo,
  S.~Panzeri, and A.~Brovelli.
\newblock Higher-order and distributed synergistic functional interactions
  encode information gain in goal-directed learning.
\newblock \emph{bioRxiv}, pages 2024--09, 2024.

\bibitem[Coronel-Oliveros et~al.(2025)Coronel-Oliveros, Gatica, Herzog, and
  Neri]{coronel2025integrated}
C.~Coronel-Oliveros, M.~Gatica, R.~Herzog, and M.~Neri.
\newblock An integrated computational framework for diversity-sensitive
  personalized medicine.
\newblock \emph{Authorea Preprints}, 2025.

\bibitem[Cramer et~al.(2010)Cramer, Waldorp, van~der Maas, and
  Borsboom]{cramerComorbidityNetworkPerspective2010}
A.~O.~J. Cramer, L.~J. Waldorp, H.~L.~J. van~der Maas, and D.~Borsboom.
\newblock Comorbidity: {{A}} network perspective.
\newblock \emph{Behavioral and Brain Sciences}, 33\penalty0 (2-3):\penalty0
  137--150, June 2010.
\newblock ISSN 1469-1825, 0140-525X.
\newblock \doi{10.1017/S0140525X09991567}.

\bibitem[Crimmins(2015)]{crimminsLifespanHealthspanPresent2015}
E.~M. Crimmins.
\newblock Lifespan and {{Healthspan}}: {{Past}}, {{Present}}, and {{Promise}}.
\newblock \emph{The Gerontologist}, 55\penalty0 (6):\penalty0 901--911, 2015.
\newblock ISSN 0016-9013.
\newblock \doi{10.1093/geront/gnv130}.
\newblock URL \url{https://doi.org/10.1093/geront/gnv130}.

\bibitem[{Cruz-{\'A}vila} et~al.(2020){Cruz-{\'A}vila}, Vallejo,
  {Mart{\'i}nez-Garc{\'i}a}, and
  {Hern{\'a}ndez-Lemus}]{cruz-avilaComorbidityNetworksCardiovascular2020}
H.~A. {Cruz-{\'A}vila}, M.~Vallejo, M.~{Mart{\'i}nez-Garc{\'i}a}, and
  E.~{Hern{\'a}ndez-Lemus}.
\newblock Comorbidity {{Networks}} in {{Cardiovascular Diseases}}.
\newblock \emph{Frontiers in Physiology}, 11, Aug. 2020.
\newblock ISSN 1664-042X.
\newblock \doi{10.3389/fphys.2020.01009}.

\bibitem[Cvijovic and Polster(2023)]{cvijovic2023network}
M.~Cvijovic and A.~Polster.
\newblock Network medicine: facilitating a new view on complex diseases.
\newblock \emph{Frontiers in Bioinformatics}, 3:\penalty0 1163445, 2023.

\bibitem[Dervi{\'c} et~al.(2024)Dervi{\'c}, Sorger, Yang, Leutner, Kautzky,
  Thurner, Kautzky-Willer, and Klimek]{dervic2024unraveling}
E.~Dervi{\'c}, J.~Sorger, L.~Yang, M.~Leutner, A.~Kautzky, S.~Thurner,
  A.~Kautzky-Willer, and P.~Klimek.
\newblock Unraveling cradle-to-grave disease trajectories from multilayer
  comorbidity networks.
\newblock \emph{Npj Digital Medicine}, 7\penalty0 (1):\penalty0 56, 2024.
\newblock \doi{https://doi.org/10.1038/s41746-024-01015-w}.

\bibitem[Dervić et~al.(2025)Dervić, Ledebur, Thurner, and
  Klimek]{dervic_comorbidity_2025}
E.~Dervić, K.~Ledebur, S.~Thurner, and P.~Klimek.
\newblock Comorbidity {Networks} {From} {Population}-{Wide} {Health} {Data}:
  {Aggregated} {Data} of 8.{9M} {Hospital} {Patients} (1997–2014).
\newblock \emph{Scientific Data}, 12\penalty0 (1):\penalty0 215, Feb. 2025.
\newblock ISSN 2052-4463.
\newblock \doi{10.1038/s41597-025-04508-9}.
\newblock URL \url{https://www.nature.com/articles/s41597-025-04508-9}.
\newblock Publisher: Nature Publishing Group.

\bibitem[Divo et~al.(2015)Divo, Casanova, Marin, Pinto-Plata, Torres, Zulueta,
  Cabrera, Zagaceta, Sanchez-Salcedo, Berto, Davila, Alcaide, Cote, and
  Celli]{divoCOPDComorbiditiesNetwork2015}
M.~J. Divo, C.~Casanova, J.~M. Marin, V.~M. Pinto-Plata, J.~P. Torres, J.~J.
  Zulueta, C.~Cabrera, J.~Zagaceta, P.~Sanchez-Salcedo, J.~Berto, R.~B. Davila,
  A.~B. Alcaide, C.~Cote, and B.~R. Celli.
\newblock {{COPD}} comorbidities network.
\newblock \emph{European Respiratory Journal}, 46\penalty0 (3):\penalty0
  640--650, 2015.
\newblock ISSN 0903-1936, 1399-3003.
\newblock \doi{10.1183/09031936.00171614}.
\newblock URL \url{https://publications.ersnet.org/content/erj/46/3/640}.

\bibitem[{European Commission. Statistical Office of the European
  Union.}(2020)]{europeancommission.statisticalofficeoftheeuropeanunion.AgeingEuropeLooking2020}
{European Commission. Statistical Office of the European Union.}
\newblock \emph{Ageing {{Europe}}: Looking at the Lives of Older People in the
  {{EU}} : 2020 Edition.}
\newblock Publications Office, 2020.
\newblock URL \url{https://data.europa.eu/doi/10.2785/628105}.

\bibitem[Forte and van~der Horst(2019)]{forteComorbiditiesMedicalHistory2019}
J.~C. Forte and I.~C.~C. van~der Horst.
\newblock Comorbidities and medical history essential for mortality prediction
  in critically ill patients.
\newblock \emph{The Lancet Digital Health}, 1\penalty0 (2):\penalty0 e48--e49,
  2019.
\newblock ISSN 2589-7500.
\newblock \doi{10.1016/S2589-7500(19)30030-5}.
\newblock URL
  \url{https://www.thelancet.com/journals/landig/article/PIIS2589-7500(19)30030-5/fulltext}.

\bibitem[Fotouhi et~al.(2018)Fotouhi, Momeni, Riolo, and
  Buckeridge]{fotouhiStatisticalMethodsConstructing2018}
B.~Fotouhi, N.~Momeni, M.~A. Riolo, and D.~L. Buckeridge.
\newblock Statistical methods for constructing disease comorbidity networks
  from longitudinal inpatient data.
\newblock \emph{Applied Network Science}, 3\penalty0 (1):\penalty0 46, 2018.
\newblock ISSN 2364-8228.
\newblock \doi{10.1007/s41109-018-0101-4}.
\newblock URL \url{https://doi.org/10.1007/s41109-018-0101-4}.

\bibitem[Haug et~al.(2020)Haug, Deischinger, Gyimesi, Kautzky-Willer, Thurner,
  and Klimek]{haugHighriskMultimorbidityPatterns2020}
N.~Haug, C.~Deischinger, M.~Gyimesi, A.~Kautzky-Willer, S.~Thurner, and
  P.~Klimek.
\newblock High-risk multimorbidity patterns on the road to cardiovascular
  mortality.
\newblock \emph{BMC Medicine}, 18\penalty0 (1):\penalty0 44, 2020.
\newblock ISSN 1741-7015.
\newblock \doi{10.1186/s12916-020-1508-1}.
\newblock URL \url{https://doi.org/10.1186/s12916-020-1508-1}.

\bibitem[Hidalgo et~al.(2009)Hidalgo, Blumm, Barab{\'a}si, and
  Christakis]{hidalgo2009dynamic}
C.~A. Hidalgo, N.~Blumm, A.-L. Barab{\'a}si, and N.~A. Christakis.
\newblock A dynamic network approach for the study of human phenotypes.
\newblock \emph{PLoS computational biology}, 5\penalty0 (4):\penalty0 e1000353,
  2009.
\newblock \doi{10.1371/journal.pcbi.1000353}.

\bibitem[Hu et~al.(2016)Hu, Thomas, and Brunak]{huNetworkBiologyConcepts2016}
J.~X. Hu, C.~E. Thomas, and S.~Brunak.
\newblock Network biology concepts in complex disease comorbidities.
\newblock \emph{Nature Reviews Genetics}, 17\penalty0 (10):\penalty0 615--629,
  Oct. 2016.
\newblock ISSN 1471-0064.
\newblock \doi{10.1038/nrg.2016.87}.

\bibitem[Jani et~al.(2019)Jani, Hanlon, Nicholl, McQueenie, Gallacher, Lee, and
  Mair]{janiRelationshipMultimorbidityDemographic2019}
B.~D. Jani, P.~Hanlon, B.~I. Nicholl, R.~McQueenie, K.~I. Gallacher, D.~Lee,
  and F.~S. Mair.
\newblock Relationship between multimorbidity, demographic factors and
  mortality: Findings from the {{UK Biobank}} cohort.
\newblock \emph{BMC Med}, 17\penalty0 (1):\penalty0 74, 2019.
\newblock ISSN 1741-7015.
\newblock \doi{10.1186/s12916-019-1305-x}.
\newblock URL \url{https://doi.org/10.1186/s12916-019-1305-x}.

\bibitem[Koskinen et~al.(2022)Koskinen, Salmi, Loukola, Mäkelä, Sinisalo,
  Carpén, and Renkonen]{koskinenDatadrivenComorbidityAnalysis2022}
M.~Koskinen, J.~K. Salmi, A.~Loukola, M.~J. Mäkelä, J.~Sinisalo, O.~Carpén,
  and R.~Renkonen.
\newblock Data-driven comorbidity analysis of 100 common disorders reveals
  patient subgroups with differing mortality risks and laboratory correlates.
\newblock \emph{Scientific Reports}, 12\penalty0 (1):\penalty0 18492, 2022.
\newblock ISSN 2045-2322.
\newblock \doi{10.1038/s41598-022-23090-3}.
\newblock URL \url{https://www.nature.com/articles/s41598-022-23090-3}.

\bibitem[Lee and Seung(2000)]{lee2000algorithms}
D.~Lee and H.~S. Seung.
\newblock Algorithms for non-negative matrix factorization.
\newblock \emph{Advances in neural information processing systems}, 13, 2000.

\bibitem[MacQueen(1967)]{macqueen1967}
J.~MacQueen.
\newblock Some methods for classification and analysis of multivariate
  observations.
\newblock \emph{Proceedings of the Fifth Berkeley Symposium on Mathematical
  Statistics and Probability}, 1:\penalty0 281--297, 1967.

\bibitem[Marinazzo et~al.(2019)Marinazzo, Angelini, Pellicoro, and
  Stramaglia]{marinazzo2019synergy}
D.~Marinazzo, L.~Angelini, M.~Pellicoro, and S.~Stramaglia.
\newblock Synergy as a warning sign of transitions: The case of the
  two-dimensional ising model.
\newblock \emph{Physical Review E}, 99\penalty0 (4):\penalty0 040101, 2019.

\bibitem[Menotti et~al.(2001)Menotti, Mulder, Nissinen, Giampaoli, Feskens, and
  Kromhout]{menottiPrevalenceMorbidityMultimorbidity2001}
A.~Menotti, I.~Mulder, A.~Nissinen, S.~Giampaoli, E.~J.~M. Feskens, and
  D.~Kromhout.
\newblock Prevalence of morbidity and multimorbidity in elderly male
  populations and their impact on 10-year all-cause mortality: {{The FINE}}
  study ({{Finland}}, {{Italy}}, {{Netherlands}}, {{Elderly}}).
\newblock \emph{Journal of Clinical Epidemiology}, 54\penalty0 (7):\penalty0
  680--686, 2001.
\newblock ISSN 0895-4356.
\newblock \doi{10.1016/S0895-4356(00)00368-1}.
\newblock URL
  \url{https://www.sciencedirect.com/science/article/pii/S0895435600003681}.

\bibitem[Monchka et~al.(2022)Monchka, Leung, Nickel, and
  Lix]{monchka2022effect}
B.~A. Monchka, C.~K. Leung, N.~C. Nickel, and L.~M. Lix.
\newblock The effect of disease co-occurrence measurement on multimorbidity
  networks: a population-based study.
\newblock \emph{BMC Medical Research Methodology}, 22\penalty0 (1):\penalty0
  165, 2022.
\newblock \doi{https://doi.org/10.1186/s12874-022-01607-8}.

\bibitem[Naghavi et~al.(2024)Naghavi, Ong, Aali, Ababneh, Abate, Abbafati,
  Abbasgholizadeh, Abbasian, Abbasi-Kangevari, Abbastabar,
  et~al.]{naghavi2024global}
M.~Naghavi, K.~L. Ong, A.~Aali, H.~S. Ababneh, Y.~H. Abate, C.~Abbafati,
  R.~Abbasgholizadeh, M.~Abbasian, M.~Abbasi-Kangevari, H.~Abbastabar, et~al.
\newblock Global burden of 288 causes of death and life expectancy
  decomposition in 204 countries and territories and 811 subnational locations,
  1990--2021: a systematic analysis for the global burden of disease study
  2021.
\newblock \emph{The Lancet}, 403\penalty0 (10440):\penalty0 2100--2132, 2024.

\bibitem[Neri et~al.(2023)Neri, Runfola, te~Rietmolen, Sorrentino, Schon,
  Morillon, and Rabuffo]{neri2023neuronal}
M.~Neri, C.~Runfola, N.~te~Rietmolen, P.~Sorrentino, D.~Schon, B.~Morillon, and
  G.~Rabuffo.
\newblock Neuronal avalanches in naturalistic speech and music listening.
\newblock \emph{bioRxiv}, pages 2023--12, 2023.

\bibitem[Neri et~al.(2024)Neri, Vinchhi, Ferreyra, Robiglio, Ates,
  Ontivero-Ortega, Brovelli, Marinazzo, and Combrisson]{neri2024hoi}
M.~Neri, D.~Vinchhi, C.~Ferreyra, T.~Robiglio, O.~Ates, M.~Ontivero-Ortega,
  A.~Brovelli, D.~Marinazzo, and E.~Combrisson.
\newblock Hoi: A python toolbox for high-performance estimation of higher-order
  interactions from multivariate data.
\newblock \emph{Journal of Open Source Software}, 9\penalty0 (103):\penalty0
  7360, 2024.

\bibitem[Neri et~al.(2025)Neri, Brovelli, Castro, Fraisopi, Gatica, Herzog,
  Mediano, Mindlin, Petri, Bor, et~al.]{neri2025taxonomy}
M.~Neri, A.~Brovelli, S.~Castro, F.~Fraisopi, M.~Gatica, R.~Herzog, P.~A.
  Mediano, I.~Mindlin, G.~Petri, D.~Bor, et~al.
\newblock A taxonomy of neuroscientific strategies based on interaction orders.
\newblock \emph{European Journal of Neuroscience}, 61\penalty0 (3):\penalty0
  e16676, 2025.

\bibitem[Partridge et~al.(2018)Partridge, Deelen, and
  Slagboom]{partridgeFacingGlobalChallenges2018a}
L.~Partridge, J.~Deelen, and P.~E. Slagboom.
\newblock Facing up to the global challenges of ageing.
\newblock \emph{Nature}, 561\penalty0 (7721):\penalty0 45--56, Sept. 2018.
\newblock ISSN 1476-4687.
\newblock \doi{10.1038/s41586-018-0457-8}.

\bibitem[Passarelli-Araujo et~al.(2022)Passarelli-Araujo, Passarelli-Araujo,
  Urbano, and Pescim]{passarelli-araujoMachineLearningComorbidity2022}
H.~Passarelli-Araujo, H.~Passarelli-Araujo, M.~R. Urbano, and R.~R. Pescim.
\newblock Machine learning and comorbidity network analysis for hospitalized
  patients with {{COVID-19}} in a city in {{Southern Brazil}}.
\newblock \emph{Smart Health (Amst)}, 26:\penalty0 100323, 2022.
\newblock ISSN 2352-6483.
\newblock \doi{10.1016/j.smhl.2022.100323}.

\bibitem[Robiglio et~al.(2025)Robiglio, Neri, Coppes, Agostinelli, Battiston,
  Lucas, and Petri]{robiglio2025synergistic}
T.~Robiglio, M.~Neri, D.~Coppes, C.~Agostinelli, F.~Battiston, M.~Lucas, and
  G.~Petri.
\newblock Synergistic signatures of group mechanisms in higher-order systems.
\newblock \emph{Physical review letters}, 134\penalty0 (13):\penalty0 137401,
  2025.

\bibitem[Roque et~al.(2011)Roque, Jensen, Schmock, Dalgaard, Andreatta, Hansen,
  Søeby, Bredkjær, Juul, Werge, Jensen, and Brunak]{roque_using_2011}
F.~S. Roque, P.~B. Jensen, H.~Schmock, M.~Dalgaard, M.~Andreatta, T.~Hansen,
  K.~Søeby, S.~Bredkjær, A.~Juul, T.~Werge, L.~J. Jensen, and S.~Brunak.
\newblock Using electronic patient records to discover disease correlations and
  stratify patient cohorts.
\newblock \emph{PLoS computational biology}, 7\penalty0 (8):\penalty0 e1002141,
  Aug. 2011.
\newblock ISSN 1553-7358.
\newblock \doi{10.1371/journal.pcbi.1002141}.

\bibitem[Rosas et~al.(2019)Rosas, Mediano, Gastpar, and
  Jensen]{rosas2019quantifying}
F.~E. Rosas, P.~A. Mediano, M.~Gastpar, and H.~J. Jensen.
\newblock Quantifying high-order interdependencies via multivariate extensions
  of the mutual information.
\newblock \emph{Physical Review E}, 100\penalty0 (3):\penalty0 032305, 2019.

\bibitem[Rousseeuw(1987)]{rousseeuw1987}
P.~J. Rousseeuw.
\newblock Silhouettes: A graphical aid to the interpretation and validation of
  cluster analysis.
\newblock \emph{Journal of Computational and Applied Mathematics}, 20\penalty0
  (1):\penalty0 53--65, 1987.
\newblock \doi{10.1016/0377-0427(87)90125-7}.

\bibitem[Runfola et~al.(2025)Runfola, Neri, Sch{\"o}n, Morillon, Tr{\'e}buchon,
  Rabuffo, Sorrentino, and Jirsa]{runfola2025complexity}
C.~Runfola, M.~Neri, D.~Sch{\"o}n, B.~Morillon, A.~Tr{\'e}buchon, G.~Rabuffo,
  P.~Sorrentino, and V.~Jirsa.
\newblock Complexity in speech and music listening via neural manifold flows.
\newblock \emph{Network Neuroscience}, 9\penalty0 (1):\penalty0 146--158, 2025.

\bibitem[Sander et~al.(2015)Sander, Oxlund, Jespersen, Krasnik, Mortensen,
  Westendorp, and Rasmussen]{sanderChallengesHumanPopulation2015}
M.~Sander, B.~Oxlund, A.~Jespersen, A.~Krasnik, E.~L. Mortensen, R.~G.~J.
  Westendorp, and L.~J. Rasmussen.
\newblock The challenges of human population ageing.
\newblock \emph{Age and Ageing}, 44\penalty0 (2):\penalty0 185--187, 2015.
\newblock ISSN 0002-0729.
\newblock \doi{10.1093/ageing/afu189}.
\newblock URL \url{https://doi.org/10.1093/ageing/afu189}.

\bibitem[Scagliarini et~al.(2023)Scagliarini, Nuzzi, Antonacci, Faes, Rosas,
  Marinazzo, and Stramaglia]{scagliarini2023gradients}
T.~Scagliarini, D.~Nuzzi, Y.~Antonacci, L.~Faes, F.~E. Rosas, D.~Marinazzo, and
  S.~Stramaglia.
\newblock Gradients of o-information: Low-order descriptors of high-order
  dependencies.
\newblock \emph{Physical Review Research}, 5\penalty0 (1):\penalty0 013025,
  2023.

\bibitem[Siah et~al.(2022)Siah, Wong, Gupta, and
  Lo]{siahMultimorbidityMortalityData2022}
K.~W. Siah, C.~H. Wong, J.~Gupta, and A.~W. Lo.
\newblock Multimorbidity and mortality: {{A}} data science perspective.
\newblock \emph{Journal of Multimorbidity and Comorbidity}, 12:\penalty0
  26335565221105431, 2022.
\newblock ISSN 2633-5565.
\newblock \doi{10.1177/26335565221105431}.
\newblock URL \url{https://doi.org/10.1177/26335565221105431}.

\bibitem[Siggaard et~al.(2020)Siggaard, Reguant, Jørgensen, Haue, Lademann,
  Aguayo-Orozco, Hjaltelin, Jensen, Banasik, and Brunak]{siggaard_disease_2020}
T.~Siggaard, R.~Reguant, I.~F. Jørgensen, A.~D. Haue, M.~Lademann,
  A.~Aguayo-Orozco, J.~X. Hjaltelin, A.~B. Jensen, K.~Banasik, and S.~Brunak.
\newblock Disease trajectory browser for exploring temporal, population-wide
  disease progression patterns in 7.2 million {Danish} patients.
\newblock \emph{Nature Communications}, 11\penalty0 (1):\penalty0 4952, Oct.
  2020.
\newblock ISSN 2041-1723.
\newblock \doi{10.1038/s41467-020-18682-4}.
\newblock URL \url{https://www.nature.com/articles/s41467-020-18682-4}.
\newblock Publisher: Nature Publishing Group.

\bibitem[Sonawane et~al.(2019)Sonawane, Weiss, Glass, and
  Sharma]{sonawane2019network}
A.~R. Sonawane, S.~T. Weiss, K.~Glass, and A.~Sharma.
\newblock Network medicine in the age of biomedical big data.
\newblock \emph{Frontiers in Genetics}, 10:\penalty0 294, 2019.

\bibitem[Sra and Dhillon(2005)]{sra_generalized_2005}
S.~Sra and I.~Dhillon.
\newblock Generalized {Nonnegative} {Matrix} {Approximations} with {Bregman}
  {Divergences}.
\newblock In Y.~Weiss, B.~Schölkopf, and J.~Platt, editors, \emph{Advances in
  {Neural} {Information} {Processing} {Systems}}, volume~18. MIT Press, 2005.
\newblock URL
  \url{https://proceedings.neurips.cc/paper_files/paper/2005/file/d58e2f077670f4de9cd7963c857f2534-Paper.pdf}.

\bibitem[Steel et~al.(2025)Steel, Bauer-Staeb, Ford, Abbafati, Abdalla,
  Abdelkader, Abdi, Zu{\~n}iga, Abiodun, Abolhassani,
  et~al.]{steel2025changing}
N.~Steel, C.~M.~M. Bauer-Staeb, J.~A. Ford, C.~Abbafati, M.~A. Abdalla,
  A.~Abdelkader, P.~Abdi, R.~A.~A. Zu{\~n}iga, O.~O. Abiodun, H.~Abolhassani,
  et~al.
\newblock Changing life expectancy in european countries 1990--2021: a
  subanalysis of causes and risk factors from the global burden of disease
  study 2021.
\newblock \emph{The Lancet Public Health}, 10\penalty0 (3):\penalty0
  e172--e188, 2025.

\bibitem[Thane(1987)]{thaneGrowingBurdenAgeing1987}
P.~Thane.
\newblock The {{Growing Burden}} of an {{Ageing Population}}?
\newblock \emph{Journal of Public Policy}, 7\penalty0 (4):\penalty0 373--387,
  1987.
\newblock ISSN 1469-7815, 0143-814X.
\newblock \doi{10.1017/S0143814X00004566}.
\newblock URL
  \url{https://www.cambridge.org/core/journals/journal-of-public-policy/article/abs/growing-burden-of-an-ageing-population/608D04DBC8999D7106ADB088B5A5B9BD}.

\bibitem[WHO(2019)]{Icd10}
WHO.
\newblock {ICD}-10 {Version}:2019, 2019.
\newblock URL \url{https://icd.who.int/browse10/2019/en}.

\bibitem[WHO(2024)]{AgeingHealtha}
WHO.
\newblock Ageing and health.
\newblock
  \url{https://www.who.int/news-room/fact-sheets/detail/ageing-and-health},
  2024.

\bibitem[Xia et~al.(2024)Xia, Tian, Xu, Zhang, Zhang, Li, and
  Wang]{xia2024global}
X.~Xia, X.~Tian, Q.~Xu, Y.~Zhang, X.~Zhang, J.~Li, and A.~Wang.
\newblock Global trends and regional differences in mortality of cardiovascular
  disease and its impact on longevity, 1980-2021: Age-period-cohort analyses
  and life expectancy decomposition based on the global burden of disease study
  2021.
\newblock \emph{Ageing Research Reviews}, page 102597, 2024.

\bibitem[Zhao et~al.(2023)Zhao, Huepenbecker, Zhu, Rajan, Fujimoto, and
  Luo]{zhaoComorbidityNetworkAnalysis2023}
B.~Zhao, S.~Huepenbecker, G.~Zhu, S.~S. Rajan, K.~Fujimoto, and X.~Luo.
\newblock Comorbidity network analysis using graphical models for electronic
  health records.
\newblock \emph{Frontiers in Big Data}, 6, 2023.
\newblock ISSN 2624-909X.
\newblock \doi{10.3389/fdata.2023.846202}.
\newblock URL
  \url{https://www.frontiersin.org/journals/big-data/articles/10.3389/fdata.2023.846202/full}.

\end{thebibliography}





\end{document}